\documentclass[]{iopart}
\expandafter\let\csname equation*\endcsname\relax
\expandafter\let\csname endequation*\endcsname\relax
\usepackage{amsmath}
\usepackage{iopams}
\usepackage{graphicx}
\usepackage{color}
\usepackage[english]{babel}
\usepackage{citesort}
\usepackage{cite}
\usepackage{ulem}
\usepackage{upgreek}
\usepackage{soul}

\usepackage{siunitx}

\usepackage[colorlinks=true,linkcolor=blue,anchorcolor=blue,citecolor=red,filecolor=green,menucolor=blue,runcolor=blue,urlcolor=blue]{hyperref}

\usepackage[utf8]{inputenc}
\usepackage{graphicx}
\usepackage[colorinlistoftodos]{todonotes}

\usepackage{cite}
\usepackage{notes2bib}
\usepackage{tikz}
\renewcommand{\epsilon}{\varepsilon}
\newcommand{\ds}{\displaystyle}

\newcommand{\mwu }{\ensuremath{\mathrm{MU}}}

\begin{document}
\title[Anelastic effects in the Kibble balance] {Flexures for Kibble balances: Minimizing the effects of anelastic relaxation}
\author{Lorenz Keck$^{1,2}$, Stephan Schlamminger$^1$, Ren\'e Theska$^2$, Frank Seifert$^3$, Darine Haddad$^1$}
\date{September 2023}
\address{$^1$ National Institute of Standards and Technology, 20899, Gaithersburg, MD, USA}
\address{$^2$ Precision Engineering Group, Technische Universit\"at Ilmenau, 98693, Ilmenau, Germany}
\address{$^3$ University of Maryland, Joint
Quantum Institute, 20742, College Park, MD, USA}
\ead{lorenz.keck@nist.gov, stephan.schlamminger@nist.gov}
\begin{abstract}
We studied the anelastic aftereffect of a flexure being used in a Kibble balance, where the flexure is subjected to a large excursion in velocity mode after which a high-precision force comparison is performed. We investigated the effect of a constant and a sinusoidal excursion on the force comparison.  We explored theoretically and experimentally a simple erasing procedure, i.e., bending the flexure in the opposite direction for a given amplitude and time. We found that the erasing procedure reduced the time-dependent force by about \SI{30}{\percent}. The investigation was performed with an analytical model and verified experimentally with our new Kibble balance at the National Institute of Standards and Technology employing flexures made from precipitation-hardened Copper Beryllium alloy C17200. 

Our experimental determination of the modulus defect of the flexure yields $1.2\times 10^{-4}$. This result is about a factor of two higher than previously reported from experiments. We additionally found a static shift of the flexure's internal equilibrium after a change in the stress and strain state. These static shifts, although measurable, are small and deemed uncritical for our Kibble balance application at present.

During this investigation, we discovered magic flexures that promise to have very little anelastic relaxation. In these magic flexures, the mechanism causing anelastic relaxation is compensated for by properly shaping and loading a flexure with a non-constant cross-section in the region of bending.
\end{abstract}

\noindent{\it Keywords\/}:
Maxwell model, generalized Maxwell model, relaxation, Kibble balance, precision measurement, flexures
\maketitle
\ioptwocol

\section{Introduction}

Work led by T.J.~Quinn, C.C.~Speake, and colleagues at the International Bureau of Weights and Measures (BIPM) in the late 1980s and early 1990s popularized flexures for the use in weighing instruments~\cite{QSD86,Quinn1992}. The team at the BIPM showed that flexures~\cite{smith2000flexures} are superior over knife edges, the traditional choice for the main and auxiliary pivots in beam balances and mass comparators. Two planes come together in a single line to form a knife edge. However, zooming in to a cross-sectional view, the meeting point is not a sharp corner but rather a curve with a finite radius~\cite{Spe87}.  This has two negative consequences affecting the location of the rotation axis for such a pivot. (1) As the knife edge moves, it effectively rolls, subtly altering the lever-arm ratio of the beam and hence changing its sensitivity and ultimately limiting the performance of the balance. (2) The Hertzian contact between the knife and the flat and, hence, the radius of curvature of the knife, depends on the load supported by the pivot. And, as Quinn~\cite{Quinn1992} notes, anelasticity and plastic deformations as well as friction in this contact causes zero point drift of the balance.
It is believed that these effects contributed to the non-stationarity of the data obtained with the third-generation Kibble balance at the National Institute of Standards and Technology (NIST) ~\cite{Schla2014,Schla2015}.  

Replacing the central knife with a flexure prevents the problems discussed above, as the flexure can be designed to have a well-defined rotation axis. A drawback of flexures, which Quinn, Speake, and coworkers were aware of and have investigated thoroughly, is the {\it anelastic relaxation}, which is also known as the anelastic aftereffect.

The simplest model of a flexure can be thought of as a perfectly elastic spring, in parallel with a lossy spring, usually described as a spring in series with a viscous element, referred to as a dashpot, see Fig.~\ref{fig:simple:model}. 
The combination of a spring and a dashpot forms what is known as a {\it Maxwell unit}, and when this Maxwell unit is combined in parallel with an ideal spring, it constitutes the {\it Maxwell model} in material science.
A material obeying the systematic of the Maxwell model is called a {\it Maxwell solid}.  When a sudden strain is applied at $t=0$ to a Maxwell solid, a significant fraction of the stress response is immediate, but the Maxwell unit contributes an exponentially decaying stress with a time constant that is given by the ratio of viscosity to the elastic modulus of the Maxwell unit.

In mass comparators, this anelastic aftereffect can be kept small by avoiding large excursions of the flexure. This can be achieved by limiting its mobility with mechanical hard stops that prevent the balance mechanism from deflecting significantly during mass placement~\cite{Quinn1992}.

For Kibble balances, however, the story is different --- large motions cannot be avoided, because two measurement modes are required: force and velocity mode~\cite{Robinson16}. The former is similar to weighing with a mass comparator, with the difference that the weight of a test mass is compared directly to a magnetic force produced by a coil. In the latter, the current-to-force coefficient of the coil is determined by sweeping the coil through the magnetic field while simultaneously measuring induced voltage and the coil's velocity. 

While it is not necessary to use the same mechanism for moving the coil and weighing the mass, it is advantageous to do so. This observation was made by Kibble and Robinson~\cite{Kibble2014} and shall be abbreviated as {\it Kibble-Robinson theory}  (KRT). 

Interestingly, the first two Kibble balances ever built~\cite{npl1,steiner2005details}, one at the National Physical Laboratory (NPL) in the United Kingdom, the other at the National Institute of Standards and Technology in the United States of America, followed the fundamental principles of KRT, i.e., to use the same mechanism for weighing and moving before the principle was known as such. The balances at NPL and NIST  used a beam and a wheel mechanism, respectively. Both were supported by a central knife-edge since large motions were necessary. In both experiments, the travel of the coil aligned with the axis of local gravitational acceleration was several centimeters, requiring rotation about the central pivot of several degrees, and flexures were not deemed suitable for this application.

Although flexure-based weighing cells have been used in Kibble balances, most of these designs ignore KRT, because a different mechanism is used to move the coil than to perform the precise weighing. Examples include the Kibble balances built in Switzerland, Korea, and at the BIPM,~\cite{FBL+20, Kim+20, EBM+22}. Not taking advantage of KRT may come at a performance cost --- a balance that outperforms the original NPL balance designed by Kibble and Robinson still needs to be built. Additionally, a larger measurement uncertainty is not the only penalty for these designs. The requirement for multiple mechanisms increases the mechanical complexity and escalates the engineering, manufacturing, and operating costs.

A flexure-based Kibble balance that obeys KRT is the panacea for this field. Hence,  it's not surprising that there is renewed interest in this topic. Researchers at the Physikalische Technische Bundesanstalt~\cite{Vas+21}, NPL~\cite{BWR18}, and NIST~\cite{Keck2022} are actively working on such systems, and the BIPM Kibble balance~\cite{BIPMprivate} is switching to a single flexure mechanism for moving and weighing, as well. Hence, it is time to revisit the anelastic relaxation of flexures and the applicability and limitations in this use case.

After exercising the mechanism for the large deflections required for velocity modes, anelastic drift from the flexures~\cite{Quinn1992} limits the accuracy obtainable from weighing in typical ABA weighing schemes (A= mass off, B= mass on). Ideally, this effect needs to be characterized and compensated for to allow for accuracy in weighing and systematic studies.  An erasing or compensation procedure shall be established to provide the means for the correct operation of such mechanisms. Erasing procedures to reduce the influence of mechanical hysteresis are well-known to researchers familiar with the Kibble balance. For example, erasing procedures are needed in knife-edge balances,  where plastic deformation is the leading cause of hysteresis. The procedure is typically a sinusoidal motion with decaying amplitude and is used to reduce errors caused by disturbances during mass exchange~\cite{SLN+01, CI14}. Such procedures are time consuming ---  especially if they need to be carried out after each mass exchange. 

In the following three sections, we present fundamental investigations into the anelasticity of the flexure mechanism for the new Kibble balance at NIST~\cite{Keck2022}. The theoretical work can be seen as the relevant extension of existing models of anelasticity at low frequencies to describe the modes of operation in the Kibble balance. These models are presented in connection with experimental results from a working prototype of the new mechanism. Measurements of the relaxation force from the flexures in servo-controlled position feedback after defined disturbance are shown.

\section{Review of anelasticity in the literature}

Anelasticity and anelastic relaxation in polycrystalline metals appeared in the literature in the late 19th century, notably by Boltzmann (1874)~\cite{Boltzmann1874} and Wiechert (1893)~\cite{Wiechert1893}. Wiechert seems to be the first to employ a generalized form of relaxation model to metals to describe an anelastic solid. His model consists of an ideal spring in parallel with multiple parallel Maxwell units. This model, which shall be called \textit{generalized Maxwell model} throughout this manuscript, was later used in configurations of a finite number, $n$, of parallel Maxwell units, e.g., in works by Nowick and Berry (1972)~\cite{Nowick72} or by Beilby \textit{et al.} (1998)~\cite{BSA98} up to a continuum of relaxation stages in work by Quinn \textit{et al.} (1992) to model low-frequency anelastic relaxation of polycristalline metals~\cite{Quinn1992b}.

One hundred years after Boltzmann and Wiechert, anelasticity and the search for materials with low internal loss garnered new interest driven by precision measurements, such as beam balances for mass comparison~\cite{Spe87}, torsion balances for accurate determinations of the Newtonian gravitational constant $G$~\cite{Speake1999}, and thermal noise limited test mass suspensions for gravitational wave antennas~\cite{Sau90}. In the 1950s, Callen \textit{et al.}~\cite{Callen1951} formulated the fluctuation-dissipation theorem linking internal damping to thermal or Brownian noise in mechanical suspensions, which counts as one of the fundamental limits to the sensitivity of high-precision mechanical detectors~\cite{Sau90}. In the context of $G$, Kuroda~\cite{Kuroda95} pointed out a possible measurement bias due to the anelastic effect in the time-of-swing method.

Experiments to measure anelastic damping can be classified into dynamic and static approaches. The former observes the ringdown of a free oscillation at the resonance frequency of the mode under investigation using the logarithmic decrement or the mechanical quality factor, $Q$, to quantify the modulus defect in a material directly~\cite{QSD95, QDS+97}. Using an inverted pendulum, for example, anelastic effects in the material can be shown very clearly and investigated for various frequencies~\cite{SSD94, YGH+05}. For these experiments, other sources of damping, for example, gas pressure forces,~\cite{Dolesi2011} have to be reduced as much as possible by performing these measurements in high or ultra-high vacuum conditions.

Beilby \textit{et al.}~\cite{BSA98} presented an interesting method to characterize open-loop static relaxation based on photoelastic measurements from optical birefringence of glass ceramics under a change in stress. The anelastic aftereffect was measured in a static experiment in optical borosilicate-crown glass (BK7) and fused silica after a sharp increase in stress had been applied and released. Beilby's report also includes dynamic measurements by exciting the eigenmode.

Quinn \textit{et al.} and Speake \textit{et al.} measured anelastic properties in dynamic ringdown experiments and static aftereffect relaxation in closed-loop null measurement with particular interest in metal flexures, especially precipitation-hardened Copper Beryllium alloy C17200 as mechanical suspension for beam and torsion balances~\cite{Quinn1992b, QSD95, QDS+97, Speake1999}. Several practical aspects of their work provide important foundations for the work presented in this manuscript which is based on observations on flexures made from similar metallurgic composition and temper.  Speake and Quinn noticed that in such polycrystalline metal flexures of minimal notch thickness down to at least \SI{50}{\micro\meter}, damping results mainly  from bulk effects and not surface effects~\cite{Quinn1992b}. The latter are more signifciant for high $Q$ materials such as sapphire or glass ceramics~\cite{Gretarsson1999, Gretarsson2000} in fibers with large surface-to-volume ratio.

\section{Physics of the  Maxwell model}

\begin{figure}[tbp]
\begin{center}
\includegraphics{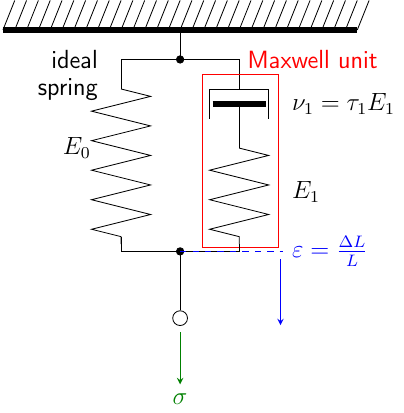}
\end{center}
\caption{The Maxwell model. An ideal spring with an elastic modulus, $E_0$, is parallel to a Maxwell unit. The Maxwell unit consists of an ideal spring with elastic modulus, $E_1$, in series with a viscous element. The viscous element known as damper or dashpot produces a velocity-dependent stress. The proportionality factor is given by the material coefficient of viscosity, $\nu_1$. The $\nu_1$ and $E_1$ quotients yield a time constant, $\tau_1$. The stress, $\sigma$, is applied in the vertical direction producing a strain, $\varepsilon$, denoting the relative length extension from the original spring length $L$.  
\label{fig:simple:model}}
\end{figure}

In this section, we review the physics of the {\it Maxwell model}, see Fig.~\ref{fig:simple:model}. The stress of the damper, $\sigma = \nu_1 \dot{\varepsilon}$, depends proportionally on the derivative of the strain $\varepsilon$ with respect to time. The proportionality constant is given by the material coefficient of viscosity, $\nu_1$. The quotient, $\nu_1/E_1$ has dimension time and is abbreviated as $\tau_1$.  For the Maxwell unit, the strain is the sum of the strain produced by the spring and the damper. This is also true for the time derivative of the strain. Hence, we obtain the following differential equation for the Maxwell unit,
\begin{equation}
\dot{\epsilon} = \frac{\dot{\sigma}_\mwu}{E_1}  + \frac{\sigma_\mwu}{\nu_1}.
\end{equation}
Performing the Laplace transformation yields,
\begin{equation}
s\breve{\epsilon} =\frac{1}{E_1} s \breve{\sigma}_\mwu +\frac{1}{\nu_1} \breve{\sigma}_\mwu.
\label{eq:laplace:mu}
\end{equation}
We use the accent $\breve{}\;$ to denote the Laplace transform of the corresponding variable.

The ideal spring has a stress-strain relationship of $\sigma_0=E_0 \epsilon$. The strain for the ideal spring is the same as for the Maxwell unit, white the stresses add, i.e.,
\begin{equation}
\sigma = \sigma_0 +\sigma_\mwu =
\frac{s \breve{\epsilon}}{\ds\frac{1}{E_1} s  +\frac{1}{\nu_1} } +E_0 \breve{\epsilon},
\end{equation}
which can be simplified to the response function of the Maxwell model
\begin{equation}
\breve{r}=\frac{\breve{\sigma}}{\breve{\epsilon}} =\frac{(E_0+E_1) s \tau_1 +E_0}{s\tau_1 +1}
\label{eq:simple:response}
\end{equation}

\subsection{Response to two simple stimuli}

\begin{figure}[htbp]
\begin{center}
\includegraphics[width=1\columnwidth]{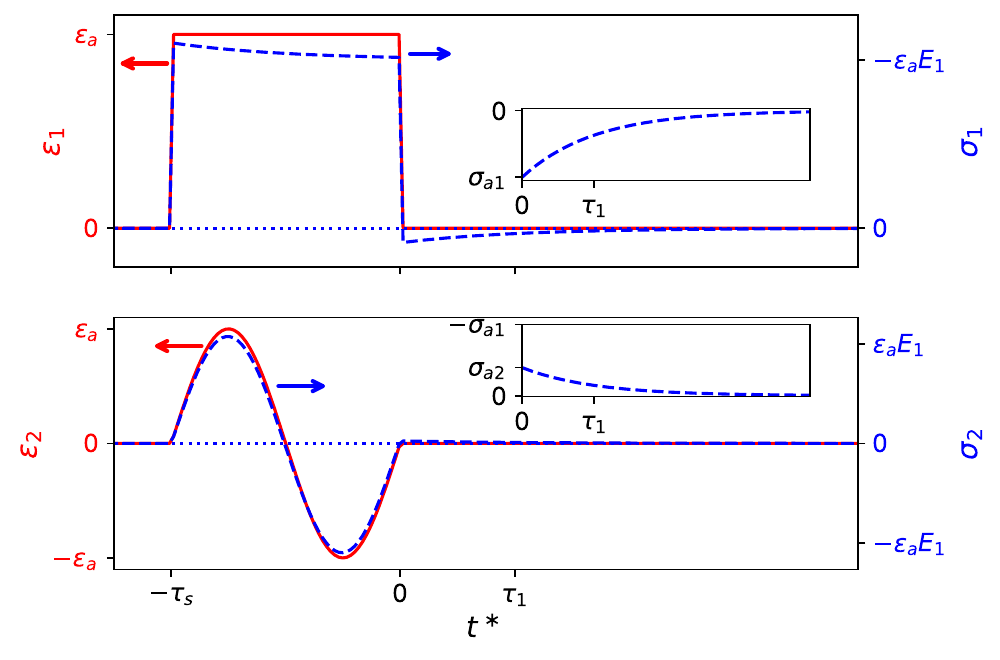}
\end{center}
\caption{Development of the stress (blue dashed line) for two strain stimuli. The top panel is for a static deflection that was held for time $\tau_S$. The bottom is a sinusoidal stimulus. At the same time, one period with identical strain amplitude was applied to the flexure. The region of interest is for $t^\ast>0$, which is shown in the insets. Note that the vertical axes of the insets have the same extent, showing that $|\sigma_{a2}|<|\sigma_{a1}|/2$. \label{fig:stain:stress}}
\end{figure}

The response function, Eq.~\ref{eq:simple:response}, allows us to calculate the time domain response, $\sigma(t)$  of any stimulus $\epsilon(t)$, by calculating the inverse Laplace transform of $\breve{r}\breve{\epsilon}$.  In the next few sections, we investigate different stimuli. The total duration of all stimuli is $\tau_s$. 
The first stimulus (subscript 1) is given in the time domain by
\begin{equation}
\epsilon_1(t) = \left\{ 
\begin{array}{ll}
0, & \mathrm{if}\; t<0\\
\epsilon_a, &   \mathrm{if}\; 0\le t<\tau_s\\
0, &  \mathrm{if}\; t\ge \tau_s\\
\end{array}
\right.
\label{eq:appl:simple:box:signal}
\end{equation}
In the Laplace domain, the above stimulus is
\begin{equation}
\breve{\epsilon}_1
=\epsilon_a  \left( \frac{1}{s} - \frac{e^{-s \tau_s}}{s}   \right).
\end{equation}
The inverse Laplace transform of the product $\breve{r}\breve{\epsilon}$ for $t\ge \tau_s$ yields the stress response. It is
\begin{equation}
\sigma_1(t) =  -\epsilon_a  E_1 e^{-t/\tau_1} \left(-1+e^{\tau_s/\tau_1}\right).
\end{equation}
The desired outcome of the calculation is the stress after a strain stimulus, which can be obtained by shifting the time axis using $t=\tau_s+t^\ast$. Then, the stress is given in the new time variable as
\begin{equation}
\sigma_1(t^\ast) = \sigma_{a1} e^{ -t^\ast/\tau_1},\;\mbox{with}\; \label{eq:decay}
\end{equation}
\begin{equation}
\sigma_{a1} 
 =-\epsilon_a  E_1\big(1-e^{ -\tau_s/\tau_1}\big)
 \label{eq:sigma:a1}
\end{equation}
consistent with Eq.~(2) in~\cite{Quinn1992b}. The imprinted stress decays with a time constant $\tau_1$. The absolute initial amplitude,  $|\sigma_{a1}|$, is larger as $\tau_s$ is longer and converges to $-\epsilon_a  E_1$ for $\tau_s\rightarrow \infty$. The top panel of Fig.~\ref{fig:stain:stress} shows the development of the stress as a function of time during and after the application of stimulus one. After the stimulus, the stress decays from a negative value to zero with a time constant, $\tau_1$. The negative value appears as the strain is suddenly released, see $t^\ast=0$ in Fig.~\ref{fig:stain:stress}.

We are not interested in a static deflection for the Kibble balance but instead, a sinusoidal deflection as the coil moves up and down in the magnetic field. To make the result comparable to the result above, we model a second (subscript 2) strain impulse as
\begin{equation}
\epsilon_2(t) = \left\{ 
\begin{array}{ll}
0, & \mathrm{if}\; t<0\\
\epsilon_a  \sin{(2 \pi  n t/\tau_s )}, &   \mathrm{if}\; 0\le t<\tau_s\\
0, &  \mathrm{if}\; t\ge \tau_s.\\
\end{array}
\right.\label{eq:mo_eps2}
\end{equation}
Here $n$ is the number of integer periods the mechanism moves in a time interval, $\tau_S$. The motion given by Eq.~\ref{eq:mo_eps2} starts and ends at $\epsilon_2=0$, which is a reasonable assumption for the Kibble experiment, where the velocity measurement is usually sandwiched between two force mode measurements for which the mechanism is controlled to the equilibrium position, $\epsilon=0$. In the $s$ domain, that impulse is
\begin{equation}
\breve{\epsilon}_2 =
\epsilon_a \frac{2\pi n/\tau_s}{s^2+4\pi^2 n^2 /\tau_s^2}   \bigg( 1-e^{-s \tau_s}\bigg).
\end{equation}
The resulting stress for $t^\ast>0$ still relaxes as $\sigma_2(t^\ast) = \sigma_{a2} e^{( -t^\ast/\tau_1)}$, but the amplitude is given by
\begin{equation}
\sigma_{a2} = \epsilon_a E_1 \frac{2 \pi n \tau_1 \tau_s}{\tau_s^2 + 4 \pi n^2 \tau_1^2} \big(1-e^{ -\tau_s/\tau_1}\big) =-\xi \sigma_{a1}.
\end{equation}
The lower panel of  Fig.~\ref{fig:stain:stress} shows the stress as a function of time for a stimulus with $n=1$. It can be seen that, for the same signal duration, the absolute value of the stress at $t^\ast$ is smaller than the one for a static deflection. The sign of the stress is also different from the constant excursion.

The attenuation of imprinted stress of a sinusoidal motion relative to a constant excursion with the same amplitude is captured by the unitless ratio $\xi$, given by the negative ratio of  $\sigma_{a2} $ to $\sigma_{a1}$, which computes to
\begin{equation}
    \xi = \frac{2\pi n \tau_1 \tau_s}{\tau_s^2+4\pi^2 n^2 \tau_1^2}.
    \label{eq:xi}
\end{equation}
We note that
\begin{equation}
\xi\le \frac{1}{2},
\end{equation}
which makes intuitive sense, because only half the time is spent straining the flexure to either side. The maximum, $\xi=1/2$ is reached at $\tau_s=2\pi n \tau_1$.

\begin{figure}[htbp]
\begin{center}
\includegraphics[width=0.95\columnwidth]{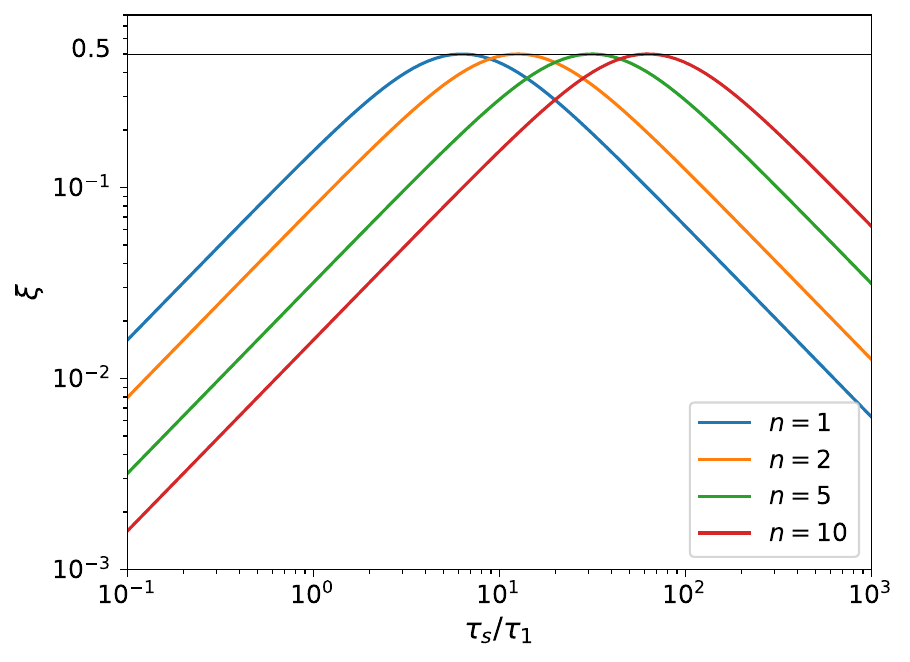}
\end{center}
\caption{The attenuation factor, $\xi$, shows the ratio of the stress amplitude after a sinusoidal motion versus a constant excursion of the same amplitude. The value $\xi$ is plotted for different ratios of signal duration, $\tau_s$, to relaxation time, $\tau_1$. The integer $n$ shows how many complete periods are in the signal duration.
\label{fig:xi}}
\end{figure}

Figure~\ref{fig:xi} shows $\xi$ as a function of $\tau_s/\tau_1$ for different integer numbers, $n$, of periods in the signal duration, $\tau_s$.  The maxima are clearly visible. On either side of the maxima, $\xi$ drops like $\tau_s/\tau_1$ or its inverse. 

The consequence of this observation is that the sinusoidal motion attenuates the relaxation effects of the Maxwell units with  $\tau_1>>\tau_s/(2\pi n)$ and $\tau_1<<\tau_s/(2\pi n)$.  Hence, in the Kibble balance with sinusoidal coil motion, the Maxwell units of concern (largest $\xi$)  have a ratio $\tau_s/\tau_1$ of order $2\pi n$. Fortunately, as discussed in the next section, the elastic energy stored in these units can be erased most effectively. 

\subsection{The boxcar eraser}

In statistics and signal processing,
a boxcar average is a convolution of a time-domain signal with a rectangular aperture~\cite{Hieftje1972}. Similarly, the erasing procedure discussed here is a rectangular motion given by two parameters, the duration, $\tau_r$, and the amplitude of the motion, i.e., strain, $\epsilon_{r1}$.  

In general, an erasing procedure is a prescribed motion that is performed after the initial stimulus with the goal of minimizing the stress after both motions are completed.  We assume that the duration of the stimulus and erasing procedure is $\tau_s$ and $\tau_r$, respectively. For the rectangular stimulus, an erasing procedure can be found such that  $\sigma(t)=0$ for $t>\tau_s+\tau_r$. 

We define a new stimulus in the time domain, labeled $\tilde{1}$ (tilde to show the case with erasing). In the time domain, it has the following functional form:
\begin{equation}
\epsilon_{\tilde{1}}(t) = \left\{ 
\begin{array}{ll}
0, & \mathrm{if}\; t<0\\
\epsilon_a, &   \mathrm{if}\; 0\le t<\tau_s\\
\epsilon_{r1}, &  \mathrm{if}\; \tau_s \ge t < \tau_s+\tau_r \\
0, &  \mathrm{if}\;   t \ge \tau_s+\tau_r \\
\end{array}
\right.
\label{eq:stim:1e}
\end{equation}
The strain amplitude $\epsilon_{r1}$ is the erasing amplitude, which depends on the erasing time, $\tau_r$. The dependence can be found as follows. The Laplace transformation of Eq.~(\ref{eq:stim:1e}) is
\begin{equation}
\breve{\epsilon}_{\tilde{1}}
=\frac{\epsilon_a}{s}  \left( 1 - e^{-s \tau_s}  \right) +\frac{\epsilon_{r1}}{s}  \left( e^{-s \tau_s} -e^{-s (\tau_s+\tau_r)} \right).
\end{equation}

\begin{figure}[htbp]
\begin{center}
\includegraphics[width=0.95\columnwidth]{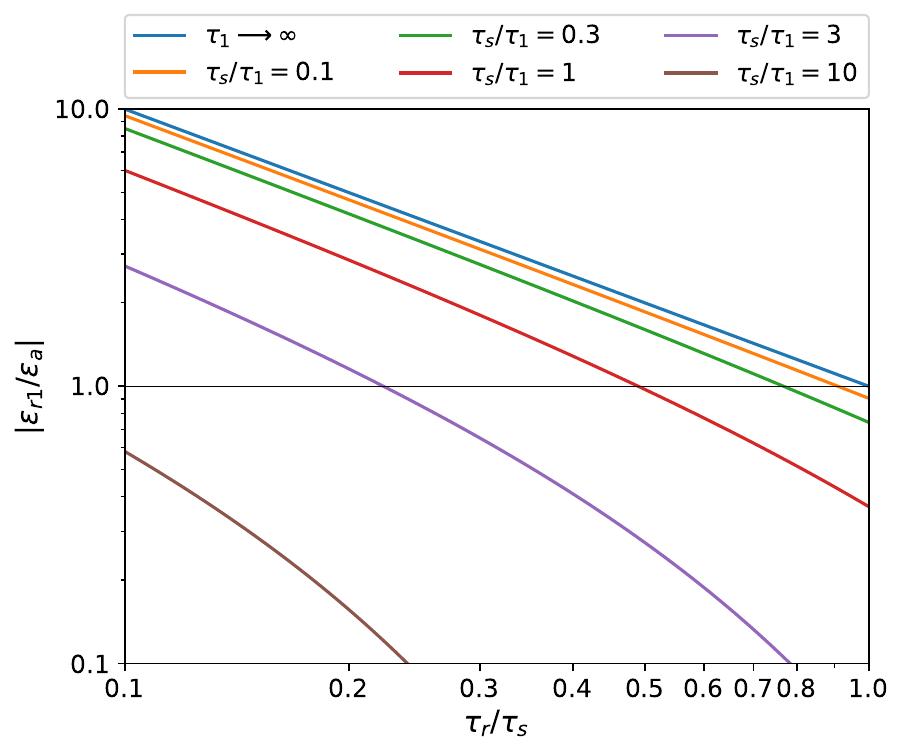}
\end{center}
\caption{The absolute value of the ratio of $\epsilon_{r1}$ to $\epsilon_a$  as a function of erasing time to stimulus time for different time constants, $\tau_1$. For a sinusoidal stimulus, the vertical axis of the graph has to be multiplied by $\xi$ given by Eq.~(\ref{eq:xi}).
\label{fig:eras:amp:time}}
\end{figure}

From the inverse Laplace transform of the product of stimulus with response function, we obtain, after shifting the time axis, this time by $t^\ast = t-\tau_s-\tau_r$ again an exponential decay, identical to Eq.~(\ref{eq:decay}), although with an amplitude
\begin{eqnarray}
\sigma_{a\tilde{1}} 
 &=&-\epsilon_a  E_1 e^{ -(\tau_s+\tau_r)/\tau_1} \big(-1+e^{ -\tau_s/\tau_1}\big) \\
 &&+\epsilon_{r1}  E_1 \big(-1+e^{ -\tau_r/\tau_1}\big).
\end{eqnarray}
This amplitude can be made zero by choosing
\begin{equation}
\epsilon_{r1} = \epsilon_a \frac{1-e^{ -\tau_s/\tau_1}}{1-e^{\ds \tau_r/\tau_1}}. \label{eq:tau:e1}
\end{equation}
The absolute value of the ratio of $\epsilon_{r1}$ to $\epsilon_a$  is shown in Fig.~\ref{fig:eras:amp:time}. In the limit $\tau_1\rightarrow \infty$, we obtain
\begin{equation}
\lim_{\tau_1\rightarrow \infty} \epsilon_{r1} = -\epsilon_a \frac{\tau_s}{\tau_r}. \label{eq:tau:e1:limi}
\end{equation}
The erasing amplitude is inversely proportional to the ratio of the erasing time to stimulus time. In other words, in this case, the relaxation can be erased with an opposite but otherwise identical signal to the stimulus. For shorter erasing procedures the erasing amplitude needs to be scaled up correspondingly.

Another special case of Eq.~(\ref{eq:tau:e1}) to consider is  $\tau_r=\tau_s$. Then,
\begin{equation}
\epsilon_{r1} = -\epsilon_a e^{ -\tau_s/\tau_1}.
\end{equation}
This equation says that the erasing stimulus needs to be opposite to the original stimulus, and the size is identical to how much the Maxwell unit has been charged after $\tau_s$. For shorter times, $\tau_r$, the absolute magnitude of the erasing amplitude becomes larger and vice versa. Experimentally, both extremes are of limited utility. (1) There is no point in using erasing times much longer than $3\tau_1$, because a similar effect can be obtained by just waiting for the decay according to Eq.~\ref{eq:decay}. (2) Short erasing times require large motion amplitudes that may not be available experimentally.

From the discussion above and Fig.~\ref{fig:eras:amp:time} it becomes apparent that it is much harder to erase Maxwell units with large time constants.

The same formalism can be applied to the sinusoidal stimulus, and not surprisingly, the results are similar, and the erasing amplitude is scaled by $-\xi$ i.e.,
\begin{equation}
\epsilon_{r2} = -\xi \epsilon_{r1} . \label{eq:tau:r2}
\end{equation}

\begin{figure}[htbp]
\begin{center}
\includegraphics[width=0.95\columnwidth]{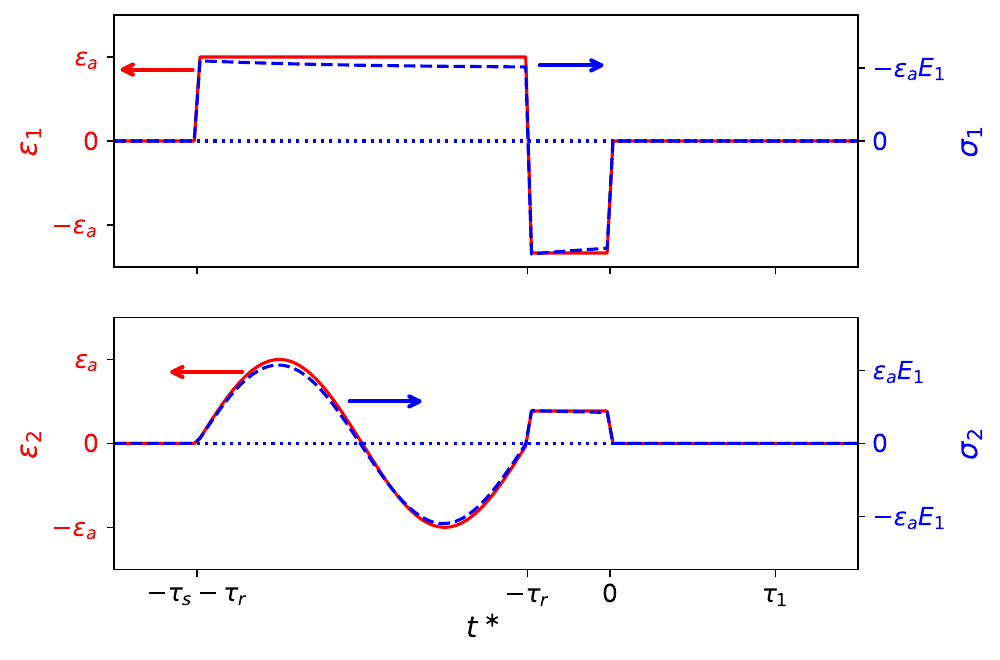}
\end{center}
\caption{Development of the stress (blue dashed line) for the same two stimuli shown in Fig.~\ref{fig:stain:stress}. This time, an erasing pulse of duration $\tau_r$ follows immediately after the stimulus. Here, the erasing pulse lasts a quarter of the original stimulus, $\tau_r=\tau_s/4$. A zoomed plot of the strain for $t^\ast>0$ is not necessary because then the stress produced by the flexure is zero.
\label{fig:stain:stress:eras}}
\end{figure}

\section{The generalized Maxwell model}

\begin{figure}[tbp]
\begin{center}
\includegraphics[width=0.95\columnwidth]{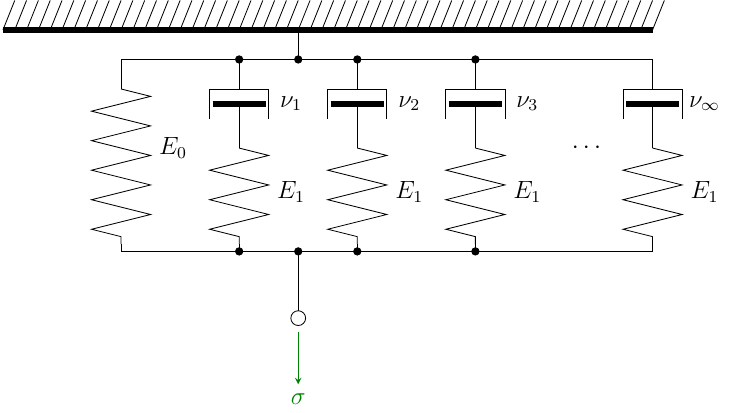}
\end{center}
\caption{The generalized Maxwell model. The main element is perfectly elastic with a modulus $E_0$. In parallel to it is an infinite number of Maxwell units, each with a spring of elastic modulus $E_1$, but different viscosity values, $\nu_j$. Hence, each unit has a different time constant $\tau_j= \nu_j/E_1$ for $j
=1,2,\cdots,\infty$. 
\label{fig:man:Maxwell:units}}
\end{figure}

Previously, we discussed the stress relaxation behavior for the Maxwell model. While this model is conducive to building intuition and deriving simple equations, it falls short of reproducing the behavior of a realistic material. A better model is an ideal spring in series with many Maxwell units having a distribution of time constants as is shown in Fig.~\ref{fig:man:Maxwell:units}. This model is known as the {\it generalized Maxwell model} and is sometimes also called the Maxwell–Wiechert model. We follow in the footsteps of Speake and Quinn, who have discussed such models in the literature~\cite{Quinn1992,Speake1999,Spe18}.

\begin{figure}[htbp]
\begin{center}
\includegraphics[width=0.95\columnwidth]{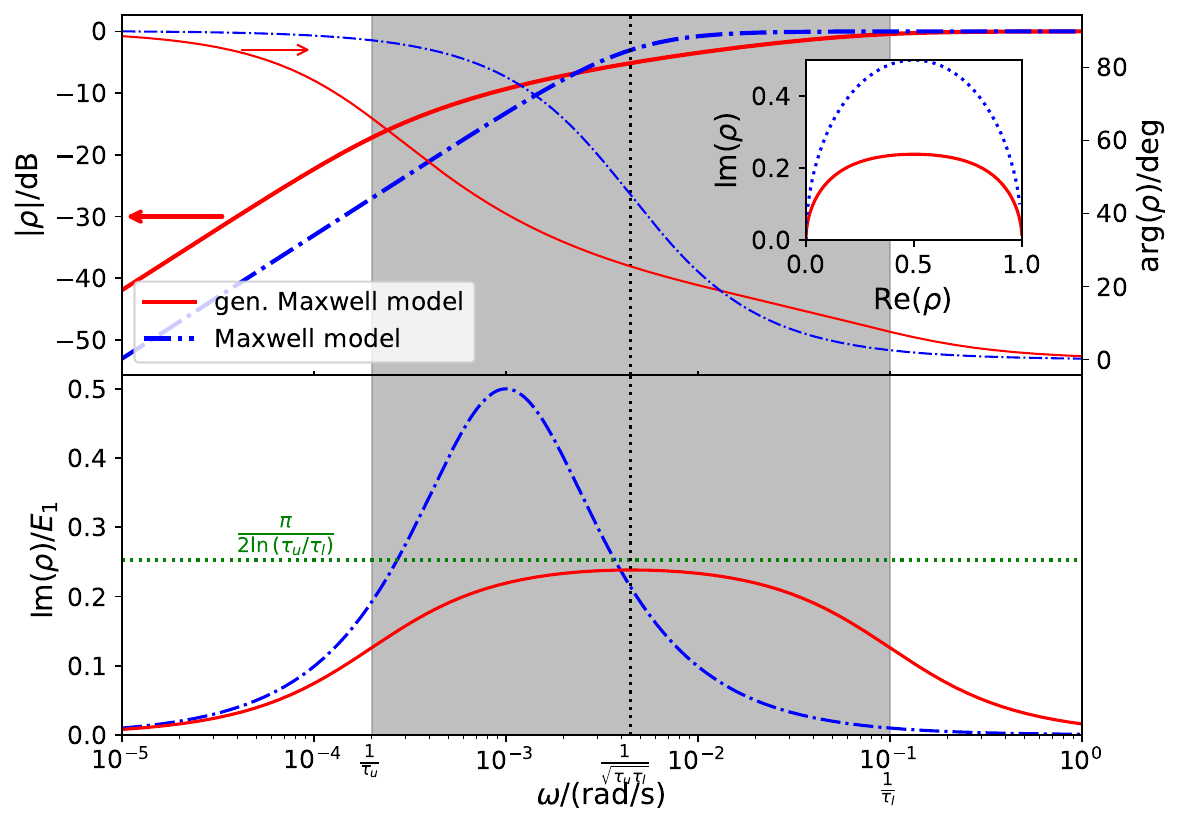}
\end{center}
\caption{ Bode plot of the anelastic part of the response function, given by $\rho:=(\frac{\breve{\sigma}}{\breve{\epsilon}}-E_0)/E_1$. The solid line gives the response function for a generalized Maxwell model with time constants distributed between $\tau_l=\SI{10}{\second}$ and $\tau_u=\SI{5000}{\second}$, see Eq.~(\ref{eq:resp:distr}).  The dashed-dotted line shows the response function for the  Maxwell model with a time constant that corresponds to the geometric mean of the limits of the distribution, i.e., $\tau_1 = \sqrt{\tau_l \tau_u}$. The inset in the top panel gives the Nyquist plot of the two responses. The response of the generalized Maxwell model is vertically compressed compared to that of the Maxwell model. The lower plot shows the imaginary part of the response normalized by $E_1$.
\label{fig:bode:distr}}
\end{figure}
First, let's assume we have $n$ discrete sets of Maxwell units, each with an elastic modulus, $E_1$, and a viscosity, $\nu_j$, yielding a time constant, $\tau_j =\nu_j/E_1$. Hence, the response function is given
by
\begin{equation}
    \frac{\breve{\sigma}}{\breve{\epsilon}} =
    E_0 + E_1 \sum_{j=1}^n \frac{s \tau_j}{s\tau_j+1}.
\end{equation}
By replacing the sum with an integral with lower limit  $\tau_l$ and upper limit $\tau_u$ and using a distribution function $f(\tau)$, we obtain
\begin{equation}
    \frac{\breve{\sigma}}{\breve{\epsilon}} =
    E_0 + E_1\int_{\tau_l}^{\tau_u} f(\tau)\frac{s \tau}{s\tau+1}\mathrm{d}\tau.
\end{equation}
Note that the distribution function is normalized such that
\begin{equation}
\int_{\tau_l}^{\tau_u} f(\tau) \mathrm{d}\tau =1.
\end{equation}
Following \cite{Quinn1992}, a good candidate is
\begin{equation}
f(\tau)  =\frac{1}{\ln{\left(\tau_u/\tau_l\right)}}  \frac{1}{\tau}.
\end{equation}

With this distribution function, the density of Maxwell units decreases for longer time constants  -- a reasonable assumption that is found in many mechanisms that produce $1/f$ noise.
The response function simplifies to
\begin{equation}
    \frac{\breve{\sigma}}{\breve{\epsilon}} =
    E_0 +E_1\frac{ \ln{\left( 1+\tau_u s\right)} -\ln{\left( 1+\tau_l s\right)}}{\ln{\left( \tau_u/\tau_l\right)}}.
\end{equation}\label{eq:resp:distr}
The second summand of the equation above is shown as a Bode Plot in Fig.~\ref{fig:bode:distr}. For comparison, the equivalent term for the Maxwell model with an effective time constant of $\sqrt{\tau_u \tau_l}$ is shown also. Qualitatively, both Bode plots look similar and show high pass behavior. However, in the region of interest, for angular frequencies between $1/\tau_u$ and $1/\tau_l$ the curves differ. More importantly, the generalized Maxwell model has about \SI{10}{\deci\bel} more gain for low frequencies.

The quotient $\frac{\breve{\sigma}}{\breve{\epsilon}}$ describes the transfer function of strain to stress ({\it the modulus}). The imaginary part is the relevant portion responsible for the anelastic effect. 
The literature~\cite{Quinn1992,Spe18} introduces the relative {\it modulus defect} as the imaginary part of the modulus of the generalized Maxwell model, which we will abbreviate with $\eta$, as
\begin{equation}
    \eta = \frac{ \ds \mathrm{Im} \left( E_1 \int_{\tau_l}^{\tau_u} f(\tau)\frac{s \tau}{s\tau+1}\mathrm{d}\tau \right) }{E_0}.
\end{equation}
Using $s= i \omega$ with $i$ being the imaginary unit, the value of the modulus defect at the geometric middle of the angular frequency range $\omega=(\tau_l\tau_u)^{-1/2}$ can be found to be
\begin{equation}
    \eta \approx \frac{\Delta E}{E_0} \;\;\mbox{with}\;\;\Delta E:=\frac{\pi}{2  \ln \left(\frac{\ds \tau_u}{\ds \tau_l}\right) } E_1 \label{eq:modulus:defect}.
\end{equation}
As is indicated in Fig.~\ref{fig:bode:distr},  the imaginary part of the anelastic response stays nearly constant between frequencies ranging from approximately  $\tau_u^{-1}$ to  $\tau_l^{-1}$. Hence, a frequency-independent modulus defect at the level given by Eq.~\ref{eq:modulus:defect} is assumed in the literature~\cite{Quinn1992b,Speake1999,Spe18} and we continue this tradition. 

Reasonable assumptions for $\tau_l$ and $\tau_u$ are \SI{10}{\second} and \SI{5000}{\second}, respectively.  With these time constants, $\pi/(2\ln{\left(\tau_u/\tau_l\right)})$ evaluates to 0.253, which is given by the green dotted line in the lower panel of Fig.~\ref{fig:bode:distr}. 

With the response function at hand, the anelastic relaxation after a rectangular strain given by Eq.~(\ref{eq:appl:simple:box:signal}) is applied can be found.
Substituting the time scale, $t=\tau_s+t^\ast$ such that $t^\ast=0$ after the applied strain, the stress is
\begin{multline}
\sigma_1(t^\ast) = \frac{\ds \epsilon_a E_1}{\ln \left(\frac{\ds \tau_l}{\ds \tau_u}\right)}\Bigg(
 \text{Ei}\left(-\frac{t^\ast}{\tau_l}\right)
 -\text{Ei}\left(-\frac{t^\ast}{\tau_u}\right) \Bigg.\\
 \Bigg.
 -\text{Ei}\left(-\frac{t^\ast+\tau_s}{\tau_l}\right)
  +\text{Ei}\left(-\frac{t^\ast+\tau_s}{\tau_u}\right)
  \Bigg),
  \label{eq:MWUdist:relax}
 \end{multline}
where $\text{Ei}$ denotes the  exponential integral given by
\begin{equation}
    \text{Ei}(x) =\int_{-\infty}^x \frac{e^t}{t} \,\mathrm{d}t.
\end{equation}
Equation~\ref{eq:MWUdist:relax} has previously been given in a slightly different form in \cite{Quinn1992b}, as Eq.~(14a).

\subsection{The boxcar eraser}
\begin{figure}[htbp]
\begin{center}
\includegraphics[width=0.95\columnwidth]{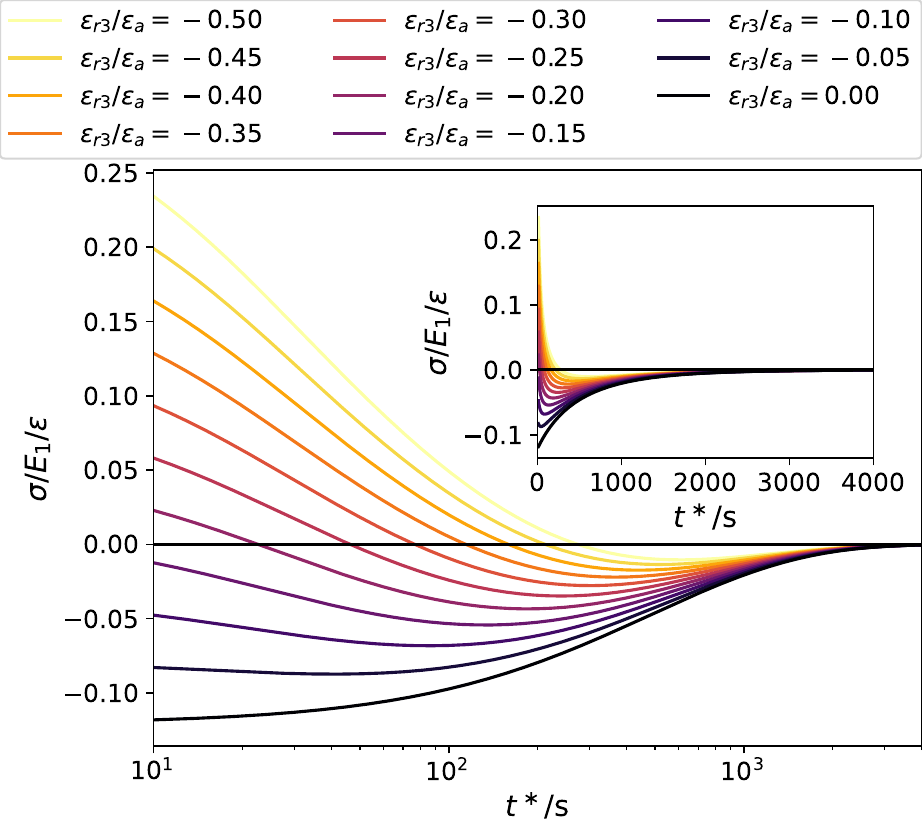}
\end{center}
\caption{Stress given by  Eq.~(\ref{eq:MWUdist:relax:after:eras}) for the generalized Maxwell model after applying the boxcar eraser. Here $\tau_r= \tau_s/4$ and $\tau_s=\SI{1920}{\second}$. The lines of different colors show different erasing amplitudes ranging from 0 to $-\epsilon_a/2$. The main plot shows the data on a logarithmic time scale. The inset shows the same data in a linear time scale.
\label{fig:box:eras:disrt}}
\end{figure}

In this section, we study the effect of the boxcar eraser given by Eq.~(\ref{eq:stim:1e}), replacing $\epsilon_{r1}$ with $\epsilon_{r3}$  for the generalized Maxwell model. In this case, the response for time $t^\ast>0$ is
\begin{multline}
\sigma_3(t^\ast) = \frac{\ds  E_1}{\ln \left(\frac{\ds \tau_l}{\ds \tau_u}\right)}\Bigg(
 \epsilon_a 
 \text{Ei}\left(-\frac{t^\ast+\tau_r}{\tau_l}\right) 
 \Bigg. \\
 -\epsilon_a \text{Ei}\left(-\frac{t^\ast+\tau_r}{\tau_u}\right)
 -\epsilon_a \text{Ei}\left(-\frac{t^\ast+\tau_r+\tau_s}{\tau_l}\right) \\
 +\epsilon_a \text{Ei}\left(-\frac{t^\ast+\tau_r+\tau_s }{\tau_u}\right)  
 +\epsilon_{r3} \text{Ei}\left(-\frac{t^\ast}{\tau_l}\right) 
  -\epsilon_{r3} \text{Ei}\left(-\frac{t^\ast}{\tau_u}\right)\\ 
 \Bigg.-\epsilon_{r3} \text{Ei}\left(-\frac{t^\ast+\tau_r}{\tau_l}\right)
 +\epsilon_{r3} \text{Ei}\left(-\frac{t^\ast+\tau_r }{\tau_u}\right) 
  \Bigg).
  \label{eq:MWUdist:relax:after:eras}
 \end{multline}
Equation~(\ref{eq:MWUdist:relax:after:eras}) is plotted for different ratios $\epsilon_{r3}/\epsilon_a$ and for the specific values of $\tau_r= \tau_s/4$ and $\tau_s=\SI{1920}{\second}$ in Fig.~\ref{fig:box:eras:disrt}. The boxcar eraser does not completely remove the anelastic relaxation. It is also interesting to note that for long $t^\ast$, i.e., $t^\ast>\tau_s$, the remaining stress is independent of the erasing amplitude and is dominated by the initial excursion.

Since perfect erasing is not attainable, we aim to find the best possible erasing procedure. To measure the effectiveness of the boxcar eraser we define a figure of merit,
\begin{equation}
f_\mathrm{om} = \int_{0}^{\infty} \big(\sigma(t^\ast)\big)^2 \,\mathrm{d}t^\ast.
\label{eq:fom}
\end{equation}
Our goal is to
 minimize $f_\mathrm{om}$.

\begin{figure}[htbp]
\begin{center}
\includegraphics[width=0.95\columnwidth]{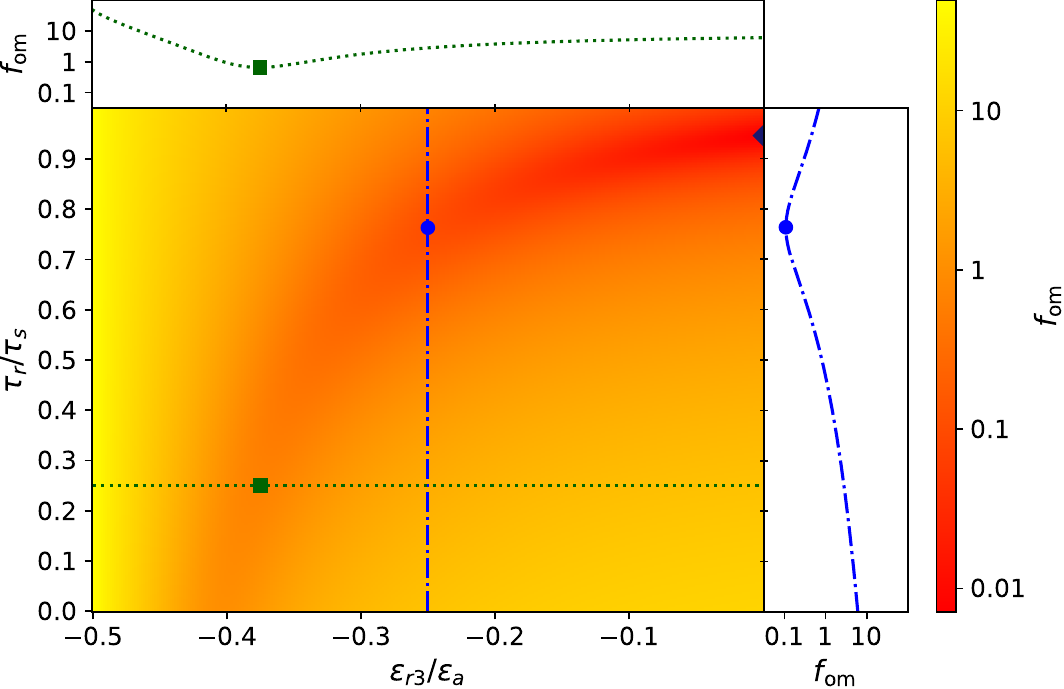}
\end{center}
\caption{The figure of merit defined by Eq.~\ref{eq:fom} as a color coded plot versus $\epsilon_{r3}/\epsilon_{a} $ on the horizontal axis and $\tau_r/\tau_s$ on the vertical axis. The plot on the right gives the figure of merit as a function of $\tau_r/\tau_s$  for $\epsilon_{r3}/\epsilon_{a}-0.25$.  The plot on the top shows the figure of merit as a function of $\epsilon_{r3}/\epsilon_{a} $ for  $\tau_r=\tau_s/4$. The square and circular markers denote the minimum along these sectional lines. The global minimum occurs at  $\tau_r=0.95/\tau_s$  and $\epsilon_{r3}=0$. It is marked with a black diamond. 
\label{fig:box:eras:FOM}}
\end{figure}

Figure~\ref{fig:box:eras:FOM} shows a heatmap of $f_\mathrm{om}$ as a function of $\tau_r/\tau_s$  and $\epsilon_{r3}/\epsilon_{a}$.  One can see a curved valley. Two points are marked in the valley. The plots to the top and the right of the heat map show the figure of merit along the horizontal and vertical dashed lines. Both marks are at the minimum along the respective dimensions. Note that the minimum marked with the blue circle is much lower than that marked with the green square. The diamond on the top right marks the point where $\epsilon_{r3}=0$. This point is equivalent to no erasing and only entails waiting.

\begin{figure}[htbp]
\begin{center}
\includegraphics[width=0.95\columnwidth]{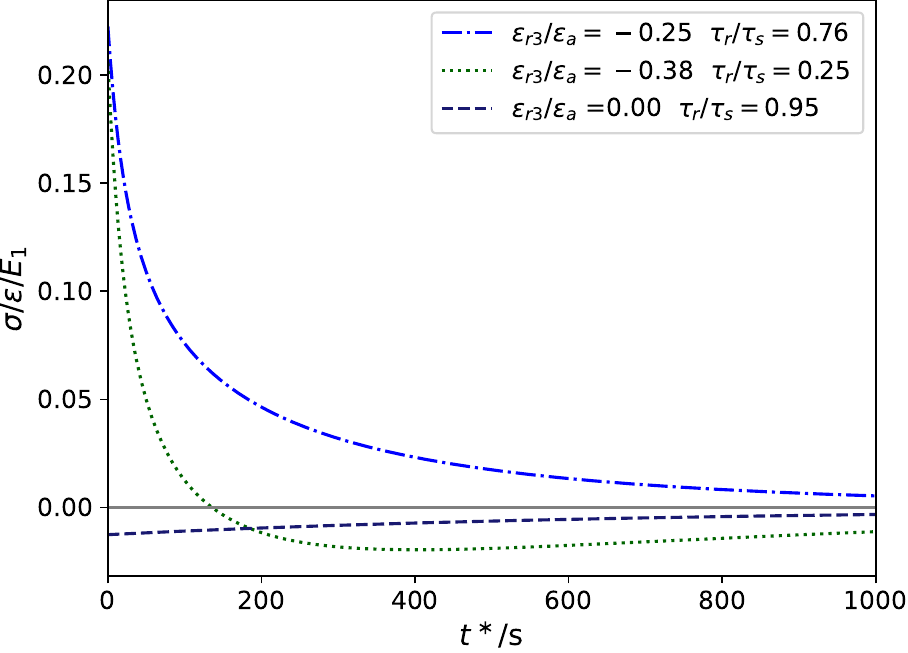}
\end{center}
\caption{The stress after the erasing procedure using the boxcar eraser. The curves correspond to the three marked parameters in Fig.~\ref{fig:box:eras:FOM}.
 \label{fig:best:box:eraser}}
\end{figure}

The stress as a function of $t^\ast$ for the three locations in the heatmap is shown in Fig.~\ref{fig:best:box:eraser}.
While simply waiting achieves the best figure of merit, it also takes the longest time, almost the same as the original excursion. The box car eraser can hasten the process. In just about a quarter of the time, a reasonable reduction in the figure of merit can be achieved.

\newcommand{\kel}{\ensuremath{\kappa_\mathrm{el}}}
\section{Effect on the measurement}

Previously, we calculated the drift in the stress produced by a linear flexure after it has been exercised sinusoidally. Here, we would like to apply this knowledge to a Kibble balance, and in order to achieve this goal, three layers of mechanics need to be revisited. (1) The basic equilibrium mechanics of the flexure-based balance. (2) The calculation of the stiffness of a flexural element. (3) The effect of one or more Maxwell units on the stiffness. Most of the details discussed below can be found in the literature~\cite{QSD95,Spe18}.  For convenience, we reiterate the main finding using coherent notation with the rest of the article. 

\subsection{The static equilibrium of the flexure-based balance}

\begin{figure}[tbp]
\begin{center}
\includegraphics[]{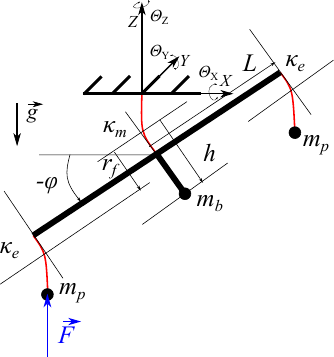}
\caption{A diagram of an equal arm balance beam suspended by a main flexure with the local gravitational acceleration, $\vec{g}$, pointing in the negative $z$ direction. All flexures are shown in red. To simplify the model, the end flexures are assumed to deflect by $\varphi$, which is a good assumption for thin flexures.\label{fig:beambal}}
\end{center}
\end{figure}
Figure~\ref{fig:beambal} shows a schematic diagram of a beam balance.  A balance beam with a total arm length of $L$ (\SI{250}{\milli\meter} in our case) and a mass, $m_b$,  is suspended from a main flexure with torsional stiffness $\kappa_m$. The beam is terminated with end flexures,  each with a torsional stiffness of $\kappa_e$ supporting a mass, $m_p$. The parameter $r_f$ denotes the distance between the line defined by the rotation axes of both end flexures and the main flexure's rotation axis.
 The force (blue arrow) on the left side of the beam balance can be controlled to maintain the right tip of the beam to an operator-defined position $z=L \sin\varphi\approx L\varphi$.

The potential energy of the system in the small-angle approximation is
\begin{equation}
    V \approx V_0+\frac{1}{2}m_b g h \varphi^2 +  m_p g r_f\varphi^2 + \frac{1}{2} \kappa_m \varphi^2+  \kappa_e\varphi^2.
\end{equation}
The restoring force imposed at an end flexure as a function of $z$ is the negative derivative of the energy with respect to $z$. It is $F=-\partial V/\partial z=-\partial V/\partial \varphi \cdot \partial \varphi/\partial z$, with $\partial \varphi/\partial z \approx 1/L$. Hence, \begin{equation}
    F\approx -\frac{m_b g h}{L} \varphi - 2 \frac{m_p g r_f}{L}\varphi -  \frac{\kappa_m}{L} \varphi-  2\frac{\kappa_e}{ L} \varphi.
\end{equation} 
By taking a second derivative with respect to $z$,  the {\it linear spring constant} can be calculated via $K=-\partial F/\partial z=-\partial F/\partial \varphi \cdot \partial \varphi/\partial z$, the spring constant can be calculated. We find
\begin{equation}
K \approx 
    \underbrace{\frac{m_b g h}{L^2}  + 2\frac{ m_p g r_f}{L^2}}_{\mathrm{gravitational}} +\underbrace{ \frac{\kappa_m}{ L^2}+  2\frac{\kappa_e} {L^2}}_{\mathrm{flexural}}.
    \label{eq:linear:spring:const}
\end{equation}
The spring constant has two principal contributions: a gravitational component, given by the first two terms, and a flexural component, given by the last two terms. Only the latter suffers from loss and anelasticity.

\subsection{Flexural stiffness}

\begin{figure}[htbp]
\begin{center}
\includegraphics[]{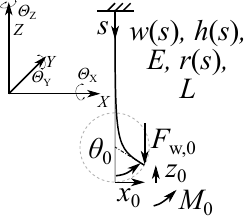}
\caption{Parameters of a flexure with load suspended in tension at the free end. The local gravitational acceleration, $\vec{g}$,  points in the negative $z$ direction. The symbols with $_0$ are values at the flexure end ($s=L$).\label{fig:flexure:params}}
\end{center}
\end{figure}

Note, in the following the symbol $s$ describes the position along the neutral fiber axis of the flexural element starting at the fixed end and not the frequency in the Laplace domain. The use of the symbol $s$ is typical for these bending problems.

\begin{table}[ht!]
    \caption{The characteristic dimensions for the elliptical notch contour of the main flexure (shown) and the circular contour for the end flexures the beam balance.\label{tab:flexure:contours} A coutour of an elliptical flexure  is shown in Fig.~\ref{fig:flexure:contour}.}
    \begin{center}
    \begin{tabular}{lS[table-format=3.1]S[table-format=3.1]S[table-format=3.1]S[table-format=3.1]}
    Dimension & \multicolumn{2}{c}{Main flexure} 
    & \multicolumn{2}{c}{End flexure}\\
    & \multicolumn{1}{l}{nominal} & \multicolumn{1}{l}{meas.}
    & \multicolumn{1}{l}{nominal} & \multicolumn{1}{l}{meas.}    \\
    \hline 
    Min. notch &50\, \si{\micro\meter}&40\, \si{\micro\meter}
    & 100\, \si{\micro\meter}& 94\si{\micro\meter}\\
    Major axis&5\,\si{\milli\meter}&\multicolumn{1}{c}{n.a.}&2.8\,\si{\milli\meter} &\multicolumn{1}{c}{n.a.}\\
    Minor axis&0.7\,\si{\milli\meter}&\multicolumn{1}{c}{n.a.}&2.8\,\si{\milli\meter} &\multicolumn{1}{c}{n.a.}\\
    Width&37.2\,\si{\milli\meter} &\multicolumn{1}{c}{n.a.}&7.6\,\si{\milli\meter}&\multicolumn{1}{c}{n.a.}
    \end{tabular}
    \end{center}
\end{table}

Here, the angular stiffness, $\kappa$, of a loaded flexure as a function of load, geometry, and elastic modulus, which is assumed to be constant, is calculated.
Following the approach using nonlinear equations of large deflections in bending used in~\cite{Lin15}  for flexures of arbitrary shape under load in tension the following two relationships hold
\begin{eqnarray}
    \frac{\mbox{d}M(s)}{\mbox{d}s} = F_\mathrm{w,0} \sin{\big(\theta(s)\big)}  &\;\mathrm{and}\;&
    \frac{\mbox{d}\theta(s)}{\mbox{d}s} = r(s) ,\label{eq:dgl:kappa}\\
    \mbox{with} \; \;  r(s)  = \frac{M(s)}{E I_\mathrm{z}(s)},&\;
    \mbox{and}&  \;  I_\mathrm{z}(s)  = \frac{b(s) h(s)^3}{12},
\end{eqnarray}   
where the second moment of area $I_z(s)$ can vary along the flexure length parameterized by $s$. The variables used in the calculation are shown in Fig.~\ref{fig:flexure:params}.

The differential equations above can be solved for $\theta(s)$. The bending of the flexure in the $x,z$ coordinate system is obtained using
\begin{equation}\label{eq:deformation:flexure}    \frac{\mbox{d}z(s)}{\mbox{d}s} = -\cos{\big(\theta(s)\big)} \;\;\mbox{and}\;\;
\frac{\mbox{d}x(s)}{\mbox{d}s} = \sin{\big(\theta(s)\big)}.
\end{equation}

This system of ordinary partial differential equations can be solved using numerical tools, such as the shooting method~\cite{Press2007}.  We have a Python~\bibnote{We have plans to make the program publicly available at the end of 2024} code that allows us to calculate arbitrarily shaped flexures. 
The end flexures can be approximated by the same calculation by reversing the boundary conditions to the ones shown in Fig.~\ref{fig:flexure:params}, which is a valid assumption at a long pendulum length of tens of centimeters of the suspended masses, $m_p$~\cite{QSD86}.

\begin{figure}[htbp]
\begin{center}
\includegraphics[width=\columnwidth]{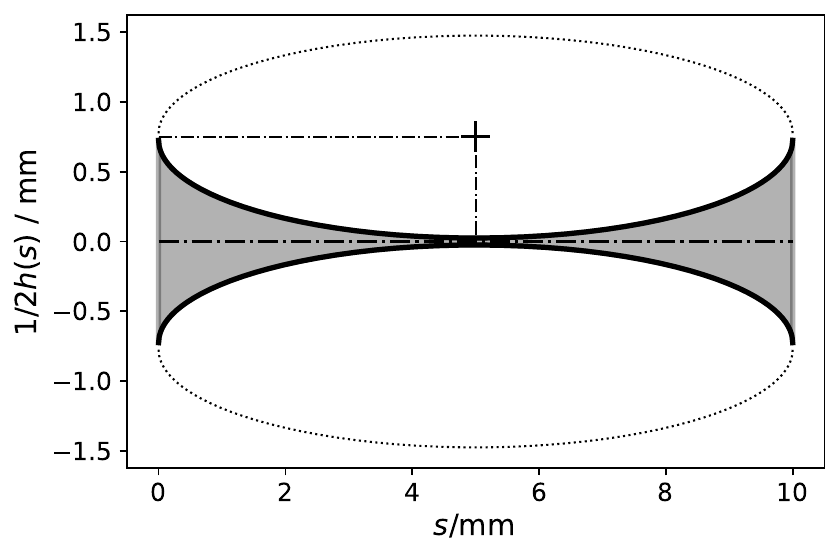}
\caption{Contour of a general elliptical flexure. The dimensions for the figure  are chosen to match our main flexure. Note the different scales in the horizontal and vertical axes of the figure. The parameter $s$ is the distance along the neutral fiber axis of the flexure. The flexure is upside-down symmetric and the parameter $h(s)$ is the thickness of the flexure. The top and bottom surfaces are given on the vertical axis. The two ellipses defining the flexure are shown with a dotted line. The minor and major axis of the upper ellipse are indicated by two dash-dotted lines. A circular contour, employed for our end flexures, is a special case in which the minor and major axis are equal.
\label{fig:flexure:contour}}
\end{center}
\end{figure}

Now, $\kappa$ can be extracted with the energy method, e.g., ~\cite{QSD95,Spe18} from the second derivative of the elastic energy with respect to the deflection angle $\theta_0$ at the free end, i.e.,
\begin{equation} 
\kappa  = \frac{\mbox{d}^2V_\mathrm{el}}{\mbox{d}\theta_0^2}
\;\;\mathrm{with}\;\;
    V_\mathrm{el} =  \int_{0}^{L} \frac{M(s)^2}{2EI_z(s)} \mbox{d}s\label{eq:Vel}.
\end{equation}

In general,  $\kappa$ can only be obtained numerically. The data describing the geometries for the two types of flexure used in the experiment are given in Tab.~\ref{tab:flexure:contours}. The numerically obtained values of $\kappa$ are shown as a function of mass load in Fig.~\ref{fig:flexure:kappa}. Additionally, $\kappa$ for the actual load are printed in Tab.~\ref{tab:flexures:stiffnesses} and a cross-sectional drawing of the flexure is shown in Fig.~\ref{fig:flexure:contour}.
Next, we need to understand the contribution of the time-dependent elastic modulus.

\subsection{Time dependent elastic modulus}
\begin{figure}[htbp]
\begin{center}
\includegraphics[width=0.95\columnwidth]{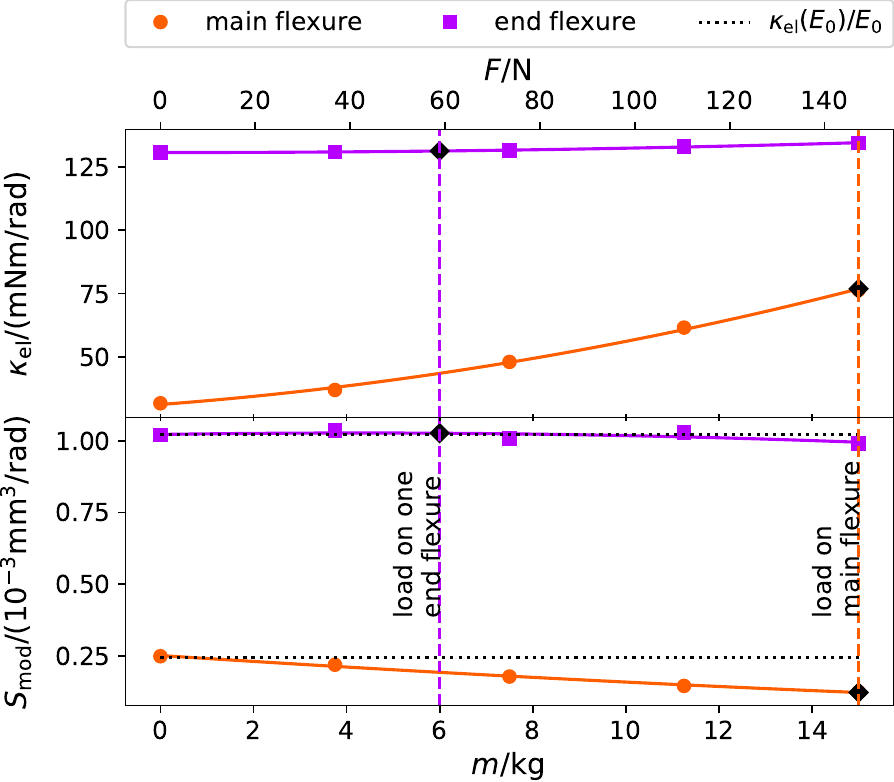}
\caption{Top plot: Results of Eq.~\ref{eq:Vel} using the solutions to numerical solutions of the coupled differential equations in Eq.~\ref{eq:dgl:kappa} as input. The points are calculated as a function of $m$, where $F=mg$. Bottom plot: 
By slightly varying the elastic modulus in the calculation shown in the top panel, the modulus sensitivity, $S_\mathrm{mod}$, can be obtained. It is plotted as a function of load. The calculation was performed for the main flexure (circles) and end flexure (squares).\label{fig:flexure:kappa}}
\end{center}
\end{figure}

The Maxwell model shows anelastic relaxation. Here, we investigate how such a change in $E$ affects  $\kappa$. Therefore, we separate our elastic modulus into time-independent and time-dependent parts,
\begin{equation}
    E = E_0 + \Delta E,
\end{equation}
where $\Delta E$ depends on time and is much smaller than $E_0$. Since $\Delta E\ll E_0$, we can Taylor expand $\kel$. It is
\begin{equation}
    \kel(E_0+\Delta E) \approx \underbrace{\kappa(E_0)}_{:=\kappa_0} +\underbrace{\left.\Delta E \frac{\mathrm{d}\kappa}{\mathrm{d}E}\right|_{E=E_0}}_{:=\Delta \kappa}.\label{eq:kappa0}
\end{equation}
Equation~\ref{eq:kappa0} also defines the static and time-dependent parts of $\kel$.
The derivative of $\kel$ with respect to $E$ can be obtained by numerically evaluating $\kel$ for discrete values of $E$ around $E_0$. Since it is an important quantity, we name it the {\it modulus sensitivity}, because it describes the sensitivity of $\kappa$ to elastic modulus. We abbreviate it with
\begin{equation}
S_\mathrm{mod} :=\left. \frac{\mathrm{d}\kappa}{\mathrm{d}E}\right|_{E=E_0}
\end{equation}
The elastic modulus for calculation of the elastic stiffness $\kappa$  was corrected from plane stress to a plane strain assumption~\cite{TSE11} by multiplication with $1/(1-\nu^2)$, where  $\nu$ is  Poisson's ratio which is 0.3 for Copper Beryllium.
The results of these calculations for the two types of flexures used in the experiment below are shown in Fig.~\ref{fig:flexure:kappa}.

For an unloaded flexure, $S_\mathrm{mod}=\kappa_0/E_0$, indicated by the dotted horizontal lines in Fig.~\ref{fig:flexure:kappa}. The higher the load, the smaller the derivative.  The relative fraction of the anelastic effect can be diluted by loading the flexure. For Kibble balance design, we advocate for the largest possible {\it anelastic dilution}, i.e., very heavily loaded flexures.

The use of dilution in precision mechanical experiments is not unheard of. Most notably, Quinn and Speake, and coworkers~\cite{Quinn2014} were using gravitational dilution by adding a lossless gravitational spring in their experiment to determine the gravitational constant $G$. More recently, Pratt and coworkers~\cite{Pratt2023} used tension in a ribbon to achieve dissipation dilution. In both of these examples, a lossless element is added in parallel to a lossy spring. The concept of the anelastic dilution discussed here is different because it minimizes $S_\mathrm{mod}$ without adding lossless restoring elements, see. Sec.~\ref{sec:magic}.

\subsection{The modulus defect}
Starting from the groundwork laid out in the three preceding sub-sections, the modulus defect can be calculated from the experimental observation. The experiments discussed below measure the force on the balance, $F$, as a function of $z$. From that, we can calculate a spring constant $K=F/z$. It is given by the gravitational and perfectly elastic terms in Eq.~\ref{eq:linear:spring:const}. The remaining time-dependent part, $\Delta K= \Delta F/z$, can easily  be converted to $\Delta \kappa=L^2 \Delta K$;  see Eq.~\ref{eq:linear:spring:const}.  According to Eq.~\ref{eq:kappa0}, $\Delta \kappa=\Delta E S_\mathrm{mod}$ and a combination of the last two equations yields,
\begin{equation}
    \eta=\frac{\Delta E}{E_0} = \frac{1}{E_0} \frac{L^2 \Delta K}{S_\mathrm{mod}} \label{eq:eta:Smod},
\end{equation}
where the numerically calculated values for the derivative are in the denominator. The numerical values for $S_\mathrm{mod}$ used here are shown in  Tab.~\ref{tab:flexures:stiffnesses}. 

In the experiment, we can only test the combination of flexures. And just as the spring constants add in Eq.~\ref{eq:linear:spring:const} so do the modulus sensitivities. As is shown in Tab.~\ref{tab:flexures:stiffnesses}, the modulus sensitivity of the whole balance 
 (last row in the table) is dominated by the end flexures. The two end flexures contribute \SI{93.8}{\percent}
to the total $S_\mathrm{mod}$. So, the experimental results given below are primarily a statement about the end flexures. The end flexures dominate because they are shorter and thicker than the main flexure, and define the region of bending more rigidly due to a larger notch effect than the main flexure, leading to negligible anelastic dilution. Small anelastic dilution values are typically not desired for Kibble balance experiments. For the investigation presented here, however, a small dilution is favorable because it emphasizes the anelastic behavior, which is the point of this article.

\begin{table}[h!]
    \caption{The characteristic stiffness and sensitivity on the elastic modulus of main flexure and end flexures for the loads given in the last column. The main flexure has a single and the end flexures a double contribution to the total.}
    \label{tab:flexures:stiffnesses}
    \centering
    \begin{tabular}{l|S[table-format=3.0]S[table-format=4.2]S[table-format=2]}
    &\multicolumn{1}{c}{$\frac{\ds \kappa}{\ds \si{\milli\newton\meter\per\radian}}$}
    &\multicolumn{1}{c}{$\frac{\ds S_\mathrm{mod}}
    {\ds \si{\milli\meter^3\per\radian}}$} 
    & \multicolumn{1}{c}{$\frac{\ds F/g}
    {\ds \si{\kilo\gram}}$} \\
    \hline 
    Main &\SI{78}{}&\SI{1.2e-4}{}&\SI{15}{}\\
    End  & \SI{130}{}&\SI{1.0e-3}{}&\SI{6}{}\\
    \hline
    Main + 2 End   & 338 & \SI{2.12e-3}{} & {n.a.} \\
    \end{tabular}
\end{table}

\subsection{Previous measurements of the modulus defect}

Several measurements of the modulus defect conducted with often hardened tempers of Copper Beryllium alloy C17200 can be found in the literature. Tab.~\ref{tab:litt:val} presents an overview of these results. Measurements were performed in bending and torsion, yielding comparable results. The earliest measurement produces a high result, which was subsequently explained by clamping losses. One study found a dependence on load.

\begin{table*}[h!]
 \caption{  Published values of the modulus defect, $\Delta E/E_0$, for CuBe flexures sorted by year of publication. \label{tab:litt:val}}
    \begin{center}
    \begin{tabular}{lp{8cm}S[table-format=4.2]}
Ref.&  Comment 
    & \multicolumn{1}{l}{$\Delta E/E_0$}\\
    \hline 
    \cite{Quinn1992}&bending, beam balance  & \SI{1.1E-3}{} \\
    \cite{QSD95} &bending and torsion. No dependence on load was found.&\SI{4.3E-5}{}\\
    \cite{QDS+97}& torsion, measured as a function of load. An update to \cite{QSD95} where no load dependence was reported&\multicolumn{1}{c}{\SI{4.3e-5}{} - \SI{6.0e-5}{}}\\
    \cite{Speake1999}&torsion, attributes the high value in\cite{Quinn1992} to losses in the flexure mount and affirms this value also for bending& \SI{5e-5}{}\\
    \cite{YGH+05}&bending, inverted pendulum  & \SI{5E-5}{} \\
    this work & see experimental section &  \SI{1.2e-4}{} 
    \end{tabular}
    \end{center}
\end{table*}

\section{Magic flexures}
\label{sec:magic}

For the flexures investigated here, the modulus sensitivity, $S_\mathrm{mod}=\mathrm{d}\kappa/\mathrm{d}E$, decreases with increasing load as is shown in the lower panel of Fig.~\ref{fig:flexure:kappa}. The question that remains is whether it is possible to build a flexure whose modulus sensitivity is zero. We call such flexures {\it magic} because they should have very little anelastic relaxation. Flexures that do not reach  $S_\mathrm{mod}=0$ but have a minimum as a function of the suspended load $F_\mathrm{w,0}$, we call {\it almost magic}.

\begin{figure}[htbp]
\begin{center}
\includegraphics[width=0.95\columnwidth]{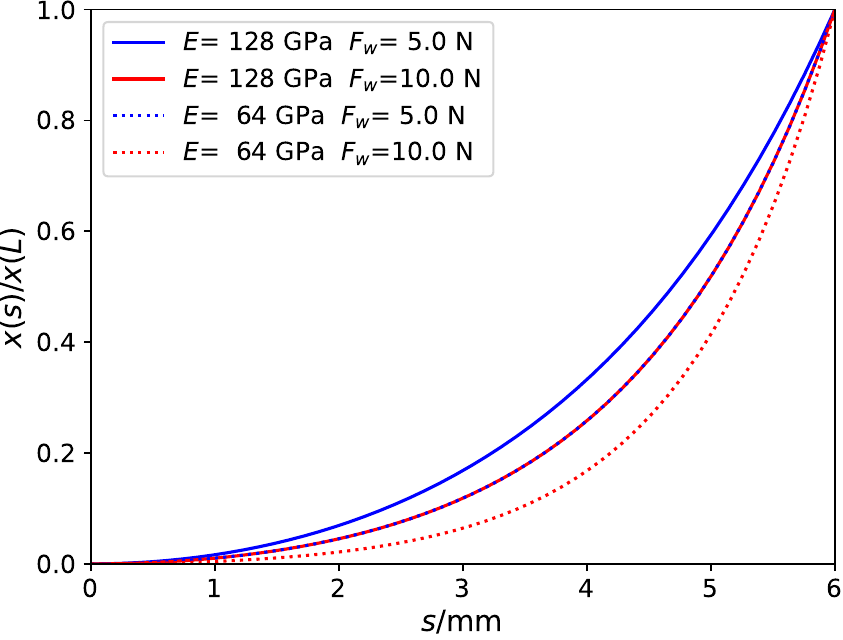}
\caption{Bending of a flat flexure under a torque of \SI{1}{\milli\newton\meter}. The dimensions of the flexure were arbitrarily chosen to be $L=\SI{6}{\milli\meter}$, $h=\SI{50}{\micro\meter}$, and $b=\SI{20}{\milli\meter}$ . The analytical solution to the differential equation for flexures of constant cross-section can be found, 
 e.g. in \cite{Spe18}. To emphasize the region where the flexure bends, the curves are normalized to the values $x(L)$, which are different for the four cases. They are in order of the legend \SI{0.17}{\micro\meter}, \SI{0.09}{\micro\meter}, \SI{0.19}{\micro\meter}, and \SI{0.10}{\micro\meter}. 
\label{fig:flat:bending} }
\end{center}
\end{figure}

While Fig.~\ref{fig:flexure:kappa} is the result of a numerical solution of the differential equation Eq.~\ref{eq:dgl:kappa}, we strive to intuitively understand the mechanism that makes certain flexures magic. To this end, we study the bending of a flat flexure of length $L$ with constant cross section $b\times h$ under applied torque. The $x$ coordinate of the flexure normalized by $x(L)$ as a function of $s$ is shown in Fig.~\ref{fig:flat:bending} for two load cases and for two different values of the elastic modulus. A larger elastic modulus moves the bending upward (to smaller $s$).

Imagine now that the cross-section of the flexure is engineered such that the area moment of inertia, $I_z$, gets smaller as the region of where the flexure bends moves up, in reaction to an increase of $E$. Then, the increase in $E$ can be counteracted by a decrease in $I_z$ and the total stiffness may stay the same. Clearly, a delicate balance  of these two effect is necessary to make the stiffness independent of $E$. But even if the effects do not cancel exactly, the dependence of $\kappa$ on $E$ can be made small. 

The parameter that helps to tune a flexure to its magic condition is the load $F_\mathrm{w,0}$ on the flexure. As shown in  Fig.~\ref{fig:flat:bending} for a flat flexure, changing the load has the opposite effect of changing the modulus. A larger load moves the bending down on the flexure (larger $s$). We note that the bending point in a flexure can be moved by changing the load.

While it is instructive to explain the emergence of magic on a flat flexure, magic flexure can only be achieved by engineering the moment area of inertia $I_z(s)$ to change as a function of $s$.  For these cases, however, analytical solutions to the differential equations governing the bending are either elusive or mathematically so complex that most insight is lost. 

The only way forward is a numerical calculation. Here, we use the code and mathematics described in the previous section to calculate the main flexure.

\begin{figure}[htbp]
\begin{center}
\includegraphics[width=0.95\columnwidth]{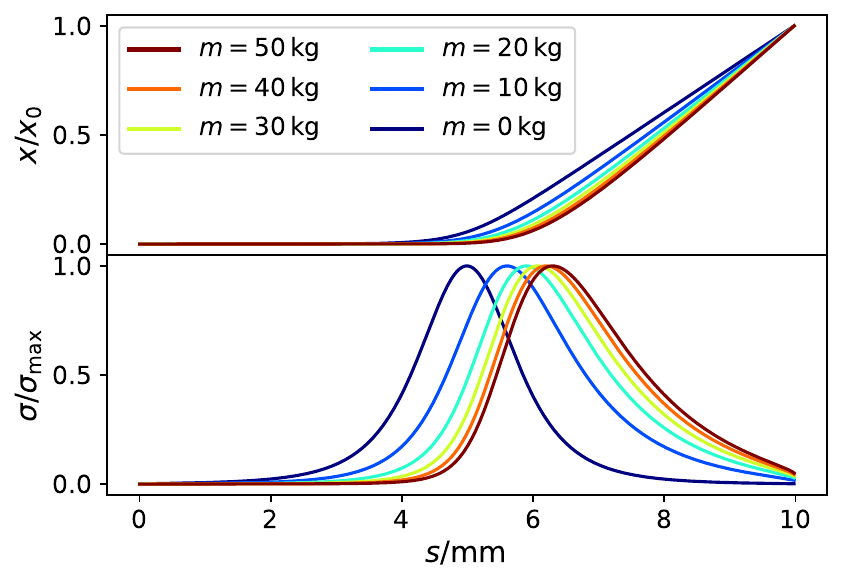}
\caption{Normalized deflection (upper graph) and normalized normal stress (lower graph) of the main flexure numerically calculated from the differential equation system in Eq.~\ref{eq:deformation:flexure} under variation of the suspended load, $F_\mathrm{w,0}=mg$. The angle $\theta_0$ was imposed as a boundary condition at the free end and the calculation was performed with  $E_0=\SI{128}{\giga\pascal}$. \label{fig:defl:main:vs:mass}}
\end{center}
\end{figure}

Fig.~\ref{fig:defl:main:vs:mass} shows the deflection as a function of the load on the flexure given by the weight, $mg$, of a point mass at the end of the flexure with mass values, $m$, ranging from \SI{0}{\kilo\gram} to \SI{50}{\kilo\gram}. Similar to to the behavior of the flat flexure shown in Fig.~\ref{fig:flat:bending}, the bending point moves down the flexure as the mass load increases. The unloaded flexure bends approximately in the middle of the flexure, $s=\SI{5}{\milli\meter}$,  where the flexure is thinnest. As the mass load increases, the bending point moves down (increase in $s$) where the flexure becomes thicker and hence stiffer.

The bending is, of course, distributed throughout the flexure. To indicate where the bending occurs the normal stress is shown in the lower panel of Fig.~\ref{fig:defl:main:vs:mass}. The maximum of the (bending) stress is a good indication of where the bending occurs. 

\begin{figure}[htbp]
\begin{center}
\includegraphics[width=0.95\columnwidth]{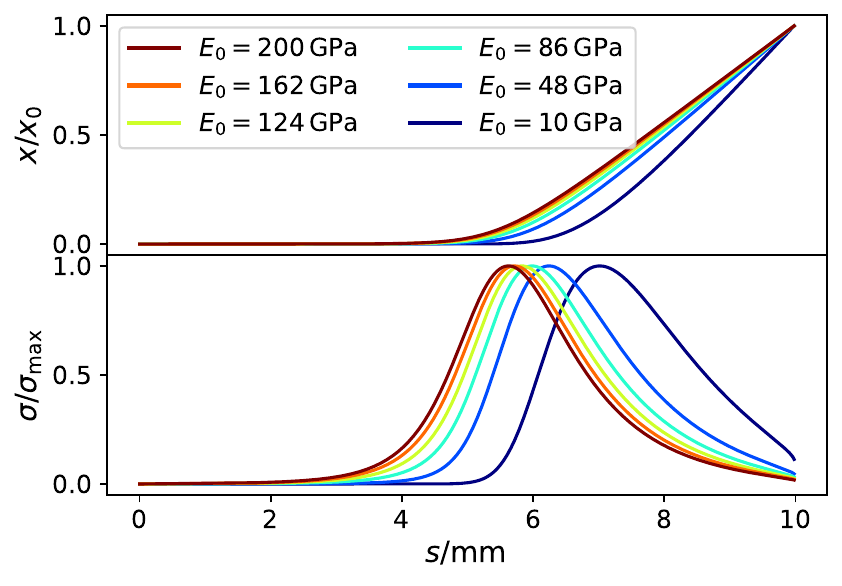}
\caption{Normalized deflection (upper graph) and normalized normal stress (lower graph) of the main flexure obtained from our Python code by changing the elastic modulus $E_0$. As before, the angle $\theta_0$ was imposed at the free end as a boundary condition. A load with $m=\SI{15}{\kilogram}$ is suspended from the flexure. \label{fig:defl:main:vs:E}}
\end{center}
\end{figure}

The complement to Fig.~\ref{fig:defl:main:vs:mass} is Fig.~\ref{fig:defl:main:vs:E} where the deflection and the stress of the same flexure are plotted for different values of $E_0$ under a constant load of \SI{15}{\kilo\gram}. We observe the opposite. As $E$ increases the bending moves towards the middle of the flexure where it is thinnest and bending happens at a lower torque. 

With this combination, it is possible to find a load where the stiffness of the flexure does not change as $E$ increases or decreases slightly, as is the case in the model of the anelastic relaxation. Usually, an increase in $E$ would increase $\kappa$, but as the location of bending moves towards a thinner part of the flexure, it simultaneously decreases. By choosing the right load, these two effects can be fine tuned such that $\kappa$ is independent of $E$, i.e., a vanishing $S_\mathrm{mod}$.

\begin{figure}[htbp]
\begin{center}
\includegraphics[width=0.95\columnwidth]{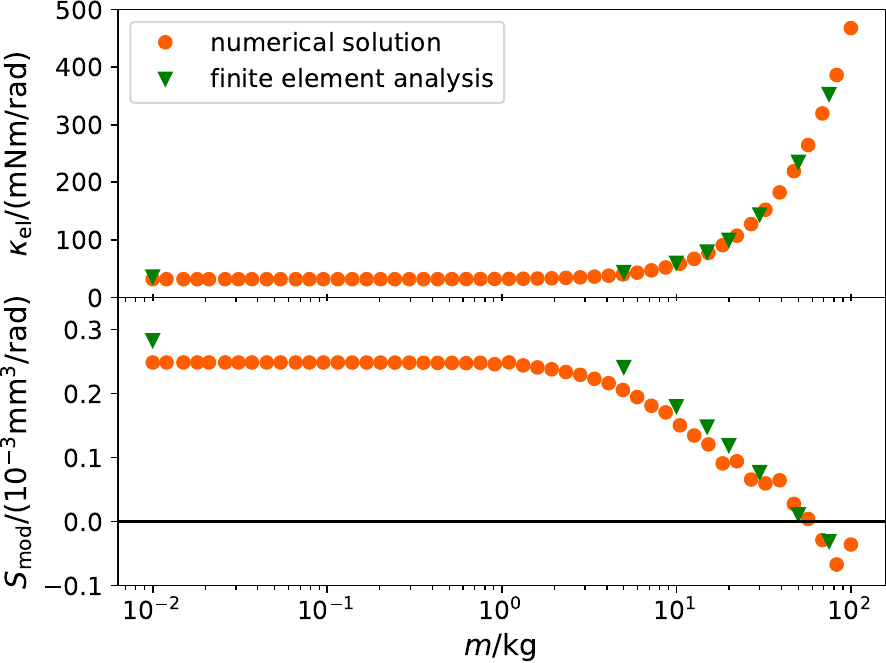}
\caption{The spring constant, $\kappa$, and the modulus sensitivity, $S_\mathrm{mod}$, as a function of load for the main flexures for a larger range of loads than in Fig.~\ref{fig:flexure:kappa}.  The orange circles are obtained by numerically solving the differential equation for the given flexure shape. The green triangles were obtained with a commercial finite element analysis software. Besides small numerical inaccuracies, the general trend agrees, and more importantly, both methods find $S_\mathrm{mod}=\SI{0}{\milli\meter^3\per \radian}$ for $m\approx \SI{56}{\kilo\gram}$. At this load in tension, the flexure has a safety factor of $3.5$ to yield, and even with a bending angle of $\theta_0=\SI{7}{^\circ}$, a safety factor of $\approx2$ is present.
\label{fig:flexure:kappa:large:mass}}
\end{center}
\end{figure}

To put this to practice, we calculate $\kappa$ and $S_\mathrm{mod}$ for the main flexure for loads up to \SI{100}{\kilo\gram}. The results are shown in Fig.~\ref{fig:flexure:kappa:large:mass}. To verify the Python code, several points were also calculated using commercial finite element analysis (FEA) software. While there are small differences between the two calculations that we attribute to negligible numerical inaccuracies in both methods, the overall trends agree. Both calculations predict the existence of a magic flexure. The main flexure becomes magic at a load of \SI{56}{\kilo\gram}. 

Much more work needs to be done to fully understand magic flexures. We believe they have low loss, i.e., high $Q$ even at low frequencies in the bandwidth of the characteristic frequencies of the generalized Maxwell solid (Fig.~\ref{fig:bode:distr}), and very little anelastic relaxation, which is our focus in this article. Hence, Kibble balances might be a perfect application for magic flexures.  The investigation presented in this section is theoretical and numerical in nature, and a careful experimental investigation is unfortunately outside the scope of this work. We hope to dive into this exciting research in years to come after the  Quantum Electro-Mechanical Metrology Suite  (QEMMS)~\cite{Keck2022} has been built and qualified.

\section{Experimental evidence}

The goal of the measurements is to understand anelastic relaxation and verify the effectiveness of the boxcar eraser. Here, we seek to arrive at a measurement sequence that reduces the effect of the anelastic relaxation on a classical ABA measurement~\cite{swanson_2010}, i.e., mass-off, mass-on, and mass-off measurement to below $\SI{2}{\nano\newton}$.
Furthermore, we would like to compare our obtained value of the modulus defect to published values. 

\subsection{Description of the apparatus}
For the experiments discussed below, the QEMMS Kibble balance was modified to only consist of a single beam, i.e., the guiding mechanism and the wheel were disconnected. With this disconnect, the system becomes much easier, and only the main flexure and the two end flexures are under investigation with this setup.

\begin{figure}[htbp]
\begin{center}
\includegraphics[width=\columnwidth]{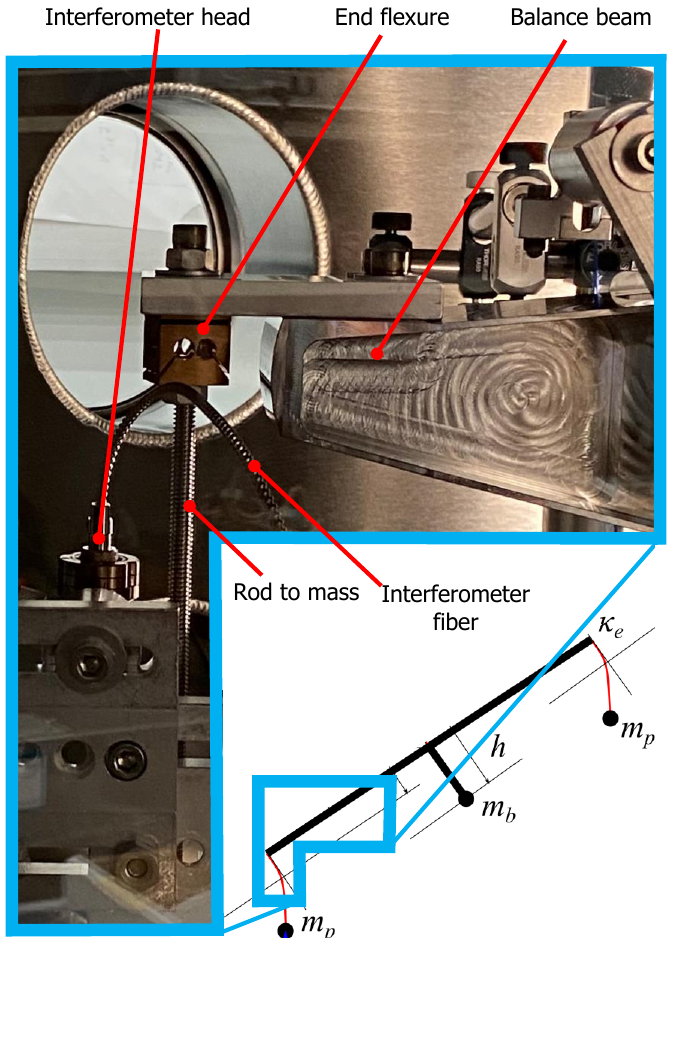}
\end{center}
\caption{Photograph of one end flexure through a window of the evacuated vacuum chamber The tapered piece of aluminum to the left is the structural part of the beam (mass $m_b$) connected to the main flexure, and the threaded rod connects to a dummy mass (mass $m_p$).
The total suspended load is $m_b+2m_p\approx \SI{15}{\kilogram}$. The top right shows a render of a monolithic end flexure with two perpendicular rotation axes, and the bottom right shows conceptually that the section of the beam balance (see Figure~\ref{fig:beambal}) framed by the red box is shown in the photograph.
\label{fig:exp:setup}}
\end{figure}

The beam balance mechanism, a position detector for feedback operation, and an electromagnetic actuator are housed in a vacuum chamber kept at \SI{0.15}{\milli\pascal}. A photo of part of the apparatus is shown in Figure~\ref{fig:exp:setup}.

This mechanism is used to suspend and guide a coil and can further be used as a high-precision force comparator. The salient aspects regarding the design are outlined in~\cite{Keck2022}. Most important for this work is that the flexure elements (main and end) are made from hardened (TF00 temper) Copper Beryllium alloy C17200 (containing about \SI{1.8}{\percent} to \SI{2}{\percent} Beryllium).  In such a modular design (Aluminum frame and beam, but Copper Beryllium flexures), it is crucial to prevent stress gradients in the interface of flexure and connecting members, which, if poorly designed, give rise to additional damping, anelastic aftereffect, and possibly also static hysteresis~\cite{QSD95}. To avoid clamping directly on flexible elements the flexures are monolithic, i.e., the thin part and the interface piece bolted onto the trusses are machined from a single coupon of material. 

The main flexure was wire cut from bulk material, which was solution heat treated and then precipitation hardened to provide a TF00 temper. It has an elliptical notch contour (see Fig.~\ref{fig:flexure:contour}) with the dimensions given in Tab.~\ref{tab:flexure:contours}. 

Each end flexure is a cube that has two pairs of perpendicular flexures intersecting at a single point. These cubes were used for the coil suspension in the NIST-4 Kibble balance~\cite{Haddad2016}.  The cube is visible in Fig.~\ref{fig:exp:setup}, it has a side length of \SI{19.05}{\milli\meter}.

\begin{figure}[htbp]
\begin{center}
\includegraphics[width=0.95\columnwidth]{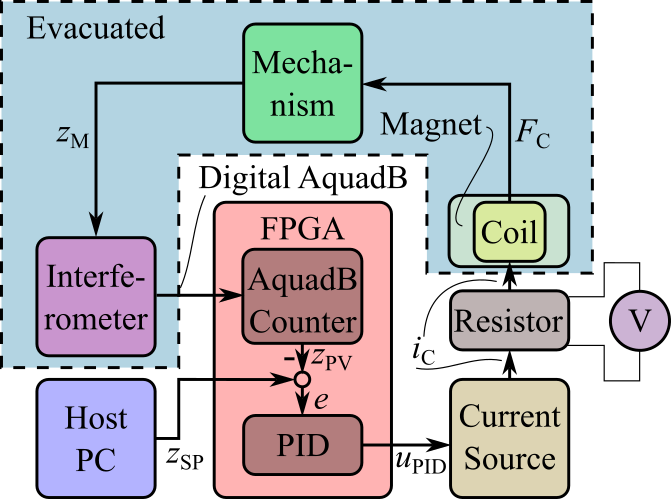}
\end{center}
\caption{Block diagram of the signals in the experiment. The host PC provides the control set point, $z_\mathrm{SP}$, to a proportional, integral, derivative (PID) controller on a field programmable gate arrays (FPGA), which compares it with the actual position reading from a quadrature encoded  (AquadB) interferometer signal. The output of the controller is a voltage, $u_\mathrm{PID}$, connected to a current source board, which directs a current, $i_\mathrm{C}$, through the coil and a calibrated resistor. A digital volt meter (V) measures current as a voltage drop over the known resistor and the coil generates a known force, $F_\mathrm{C}$, in an electromagnetic setup with a permanent magnet rigidly fixed to the experiment's frame. This force acts upon the mechanism and can keep the mechanism in feedback controlled null state. The mechanism displacement, $z_\mathrm{M}$, from the ideal position, $z_\mathrm{SP}$, is monitored.
\label{fig:schematic}}
\end{figure}

The schematic of the block diagram of the control loop is shown in Fig.~\ref{fig:schematic}. For the data reduction and analysis, two quantities are important. The position of the balance, $z_M$, and the force, $F$, is required in order to control the balance to $_\mathrm{M}$. Here,  $z_\mathrm{M}$ is the vertical position of a mass suspended from the left (in Fig.~\ref{fig:exp:setup}) end flexure. The force is obtained by multiplying the coil current, $i_\mathrm{C}$, with the geometric factor, $Bl$. The coil current is measured as a voltage drop over a calibrated resistor. The geometric factor has been calibrated to three digits by measuring a known mass. 

\subsection{Sources of measurement uncertainties}

The main sources of measurement and analysis uncertainties are summarized below:
\begin{enumerate}
    \item Fit uncertainty, mostly due to the defined start of $t^\ast$. We fit a perfect mathematical description to an imperfect experiment.

    \item     Since velocity mode was not available yet, we calibrated the $Bl$ with a standard mass, whose value was known with a relative uncertainty of \SI{1e-3}{}. We assign this number to the relative uncertainty of the force measurements. 
    \item The measurements discussed below were made with very little current in the coil as the instrument was balanced to better than \SI{300}{\micro\newton} leading to two benefits:
    \begin{itemize}
        \item  Thermal effects due to ohmic heating in the coil were negligible.
        \item Magnetic hysteresis is not a concern at these small currents, up to \SI{4}{\micro \ampere} at static holding.
    \end{itemize}
    \item Thermal effects due to evacuation and room temperature fluctuations affect the measurements at long-time scales. To account for these, a linear drift is removed from the measurements during data processing.
    \item Similarly, ground tilt can shift the center of mass and can cause long-term drift in the required holding current in the balance. Similarly to thermal effects, tilt produces a $1/f$ effect and is subtracted by linear fitting to the data.
    \item The flexure clamps are designed to act on rigid pieces of metal, never on flexing elements.
    \item When deflecting the balance mechanism from one position to another, there is a trade-off between the duration of movement and the overshoot from the commanded position due to the large inertia of the structure. The overshoot is of order $<\SI{2}{\percent}$  and lasts less than $\SI{10}{\second}$. Hence, such overshoots only trigger the very short time constants of a generalized Maxwell solid and are negligible for the long time constants that we are interested in; See Fig.~\ref{fig:xi}. 
    \item The coil current is supplied by wires attached to the balance mechanism. To minimize parasitic elastic and hysteretic effects produced by these wires, they are very thin (AWG 50 about \SI{25}{\micro\meter} in diameter, including insulation). They are routed parallel to the flexing axis of the main pivot. The wire stiffness is further reduced by using extra wire, about \SI{50}{\milli\meter}, which is coiled into a spring shape.  The wires running from the wheel to the coil are wound to springs as well, long, and do not move horizontally more than \SI{1.5}{\milli\meter} for the largest imposed displacements of the mechanism.    
\end{enumerate}

Despite the limitations discussed above, the measurement results provide valuable information for the anelastic relaxation.

\subsection{ Verifying the generalized Maxwell Model}

\begin{figure}[htbp]
\begin{center}
\includegraphics[width=0.95\columnwidth]{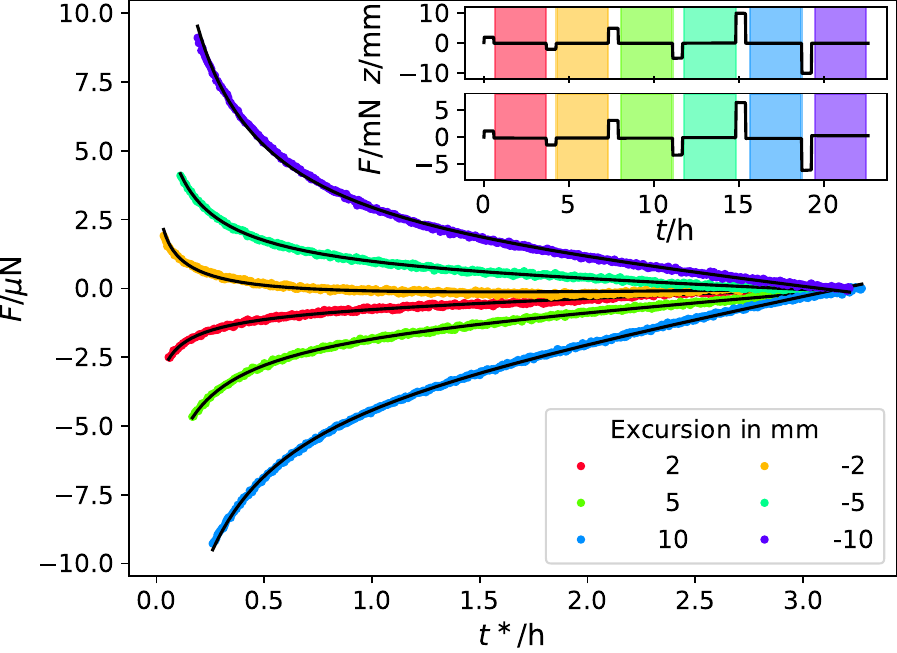}
\end{center}
\caption{The anelastic relaxation for six different excursion amplitudes measured with the simple beam balance. The measurement procedure is shown in the inset. The top graph of the inset shows the balance excursions in mm. Each excursion is held for 32 minutes. After the balance has been servoed to zero, the force required to hold the balance at zero is recorded for 3 hours, indicated by the shaded regions. The corresponding data points are plotted as the colored data points in the main plot. The black lines are fits to Eq.~(\ref{eq:fit:relax}) using $F_A$ as a prefactor.
\label{fig:measured:relaxation}}
\end{figure}

The first series of measurements is summarized in Fig.~\ref{fig:measured:relaxation}. For the day-long experiment, the balance position, $z$, was controlled to \SI{\pm 2}{\milli\meter}, \SI{\pm 5}{\milli\meter}, and \SI{\pm 10}{\milli\meter}. In each position, the beam was held for 32 minutes. After that, the force required to maintain the balance at $z=0$ was measured for three hours. 

Two steps of post-processing were applied to the data. From all measurements, a constant balance drift of \SI{-1.3}{\micro\newton \per \hour} was subtracted.  The experimental run was started within 24 hours of pumping the system to vacuum. The temperature change caused by the evacuation led to the aforementioned drift, which was numerically taken out. Usually, after a few days in the vacuum, the drift has vanished.

Second, each trace was shifted vertically such that the fit result converges to a $F=0$ for $t\rightarrow \infty$. Subtracting these offsets makes the plot less discombobulated and shows clearly that the relaxation amplitude is proportional to the initial deflection. This processing step hides one observation, though. For the large amplitudes, i.e., \SI{\pm 10}{\milli\meter}, the equilibrium position of the balance changes as much as \SI{3.5}{\micro\newton}. These irreversible changes were investigated separately and are discussed in detail in Sec.~\ref{sec:Irreversible:effects}.

\begin{table}[h!]
    \caption{Amplitude of the anelastic relaxation. The original excursion is given in the first column. The amplitude of the anelastic relaxation is labeled $F_A$. The column with the header $\eta$ gives the modulus defect. It is obtained using Eq.~(\ref{eq:mod:def}). Applying the boxcar eraser reduces the amplitude to $F_B$. The ratio of the reduction is given in the last column.\label{tab:amp:anelastic}}
    \begin{center}
    \begin{tabular}{S[table-format=4]
    S[table-format=3.1]S[table-format=8.6]S[table-format=3.1]S[table-format=4]}
    \multicolumn{1}{c}{$z_A$} 
        & \multicolumn{1}{c}{$F_A$}
    & \multicolumn{1}{c}{$\eta$}
    & \multicolumn{1}{c}{$F_B$}
    & \multicolumn{1}{c}{$F_B/F_A$}
\\   
        \multicolumn{1}{c}{(\si{\milli\meter}) }
    & \multicolumn{1}{c}{(\si{\micro\newton})}
    & \multicolumn{1}{c}{}
      & \multicolumn{1}{c}{(\si{\micro\newton})}
      & \multicolumn{1}{c}{(\si{\percent})}
      \\
        \hline 
    -10 &7.1&\SI{1.6e-4}{}& -2.5 & -36 \\
     -5&     2.3&\SI{1.1e-4}{} &-0.9 & -39\\
    -2 &     1.0&\SI{1.1e-4}{}& -0.2 & -19\\
    2&        -0.8 &\SI{9.2e-5}{}& 0.2 & -25\\
    5&      -2.5 &\SI{1.1e-4}{}& 0.8 & -32\\
    10&       -6.4&\SI{1.5e-4}{}& 2.6 & -40\\
    \hline
    \multicolumn{2}{c}{Average:}&\SI{1.2e-4}{}&&-27\rule{0pt}{2.5ex}\\
    \multicolumn{2}{c}{Std. Dev.:}&\SI{2.7e-5}{}&&8

    \end{tabular}
    \end{center}
\end{table}

A least squares fit was used to fit  the function
\begin{multline}
F(t^\ast) = F_A\Bigg(
 \text{Ei}\left(-\frac{t^\ast}{\tau_l}\right)
 -\text{Ei}\left(-\frac{t^\ast}{\tau_u}\right) \Bigg.\\
 \Bigg.
 -\text{Ei}\left(-\frac{t^\ast+\tau_s}{\tau_l}\right)
  +\text{Ei}\left(-\frac{t^\ast+\tau_s}{\tau_u}\right)
  \Bigg)
  \label{eq:fit:relax}
 \end{multline}
to the data. The functional form of this equation, with the exception of the prefactor, and hence the unit,  is identical to  Eq.~(\ref{eq:MWUdist:relax}). The only free parameter is the amplitude $F_A$, as the time constants were fixed to be $\tau_l=\SI{10}{\second}$ and $\tau_u=\SI{5000}{\second}$.
The time constants were fixed to the values populated in literature, e.g., in~\cite{Quinn1992, Quinn1992b} to avoid getting a randomly better fit without physical meaning from a combination of unphysical values.

From $F_A$ the modulus defect, $\eta$, can be obtained by combining Eq.~(\ref{eq:eta:Smod}) with $\Delta K= F_A/ z_a$ and is found to be
\begin{equation}
\eta =
    - \frac{1}{E_0} \frac{L^2}{S_\mathrm{mod}} \frac{F_A}{z_a},\label{eq:mod:def}
\end{equation}
where $z_a$ is the original excursion.

A close inspection of Fig.~\ref{fig:measured:relaxation} leaves us with several observations. (1) The anelastic aftereffect can take a long time. The plot shows three hours after the excursion, and the anelastic effect has not completely decayed. (2) Eq.~(\ref{eq:fit:relax}) is a reasonable fit to the measured traces. For the fits used in Fig.~\ref{fig:measured:relaxation}, $\tau_l$ and $\tau_u$ have been fixed to \SI{10}{\second} and \SI{5000}{\second}, respectively. (3) The size of the anelastic aftereffect is roughly proportional to the original excursion, consistent with Eq.~(\ref{eq:fit:relax}), which is proportional to $z_A$. Tab.~\ref{tab:amp:anelastic} gives the fitted amplitude for the four measurements.

The calculated modulus defects are shown in Tab.~\ref{tab:amp:anelastic}. While there is considerable scatter in the data, the average is $\eta=\SI{1.2e-4}{}\pm \SI{2.7e-5}{}$, where we used the sample standard deviation as an uncertainty.  This number is a factor of two to three larger than the results previously reported in the literature; see Tab.~\ref{tab:litt:val}. Given the difficult nature of precisely measuring the modulus defect, this agreement is satisfactory. Furthermore, it can be concluded that the mechanics is {\it not dominated by clamping loss}, because  damping loss would lead to an amplitude dependent modulus defect~\cite{QSD95}, which was not detected..

\subsection{ Verifying the boxcar eraser}

Here, the effectiveness of the box car eraser is tested. The experiment was conducted in the same way as previously. The balance was held for 32 minutes each at the $z$-positions,  \SI{\pm 2}{\milli\meter}, \SI{\pm 5}{\milli\meter}, and \SI{\pm 10}{\milli\meter}. After this time, the balance is controlled to the negative value of the original setpoint and left there for 8 minutes. After that, it is controlled to 0, and the force is measured. 

\begin{figure}[htbp]
\begin{center}
\includegraphics[width=0.95\columnwidth]{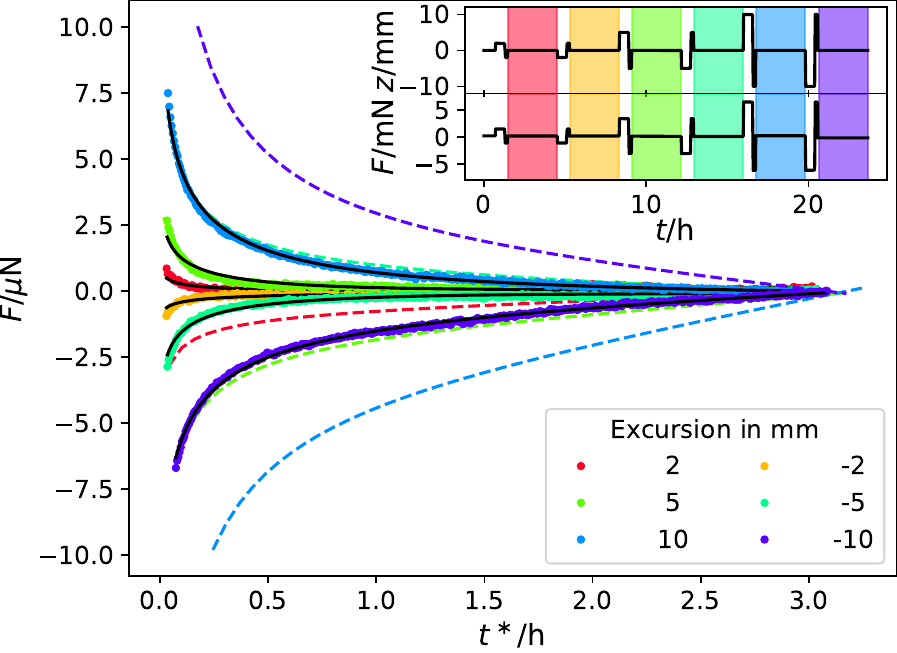}
\end{center}
\caption{The balance is stressed in equivalent manner as in Fig.~\ref{fig:measured:relaxation}. This time, after holding the original excursion for 32 minutes, the balance is controlled to the negative excursion and held there for 8 minutes, see the top graph in the inset. The colored data points show the measured force on the balance for three hours after erasing. The black lines are fits to Eq.~(\ref{eq:MWUdist:relax:after:eras}). The dashed colored lines are the fits to the data without erasing procedure, i.e., the black lines in Fig.~\ref{fig:measured:relaxation}.
\label{fig:measured:relaxation:with:eras}}
\end{figure}

The results are shown in Fig.~\ref{fig:measured:relaxation:with:eras}. To allow for easy comparison to the traces without erasing, the original fits (black lines in Fig.~\ref{fig:measured:relaxation}) are shown here as dashed lines with corresponding colors. After the boxcar eraser, the amplitudes are much smaller, and the integral between the curve and the $F=0$ line is much smaller. The new amplitudes, $F_B$ are given in Tab.~\ref{tab:amp:anelastic}. One can see that, on average, the relaxation amplitude is a bit more than a quarter of the original amplitude. Note the minus sign in the table. After erasing the flexure, the sign of the anelastic relaxation changed and the residual forces were the opposite as they would have been without the erasing procedure.

\section{Irreversible effects}
\label{sec:Irreversible:effects}
Analyzing the results of the measurements discussed in the preceding section, we noticed that, especially after large displacement, the flexure did not come back to the same zero position -- even after a long time. This effect is known and is another reason, why large excursions of flexures are generally avoided.  It is as if the spring had been reset after a large excursion. In the Kibble balance, such a reset of the spring after velocity mode is generally not a problem as long as the spring stays consistent in the subsequent mass-on and mass-off measurements.  However, the mass manipulation for the mass-on and mass-off measurements also induces disturbances to the balance, and it would be important to know if the disturbances are large enough to cause a systematic bias.

\begin{figure}[htbp]
\begin{center}
\includegraphics[width=0.95\columnwidth]{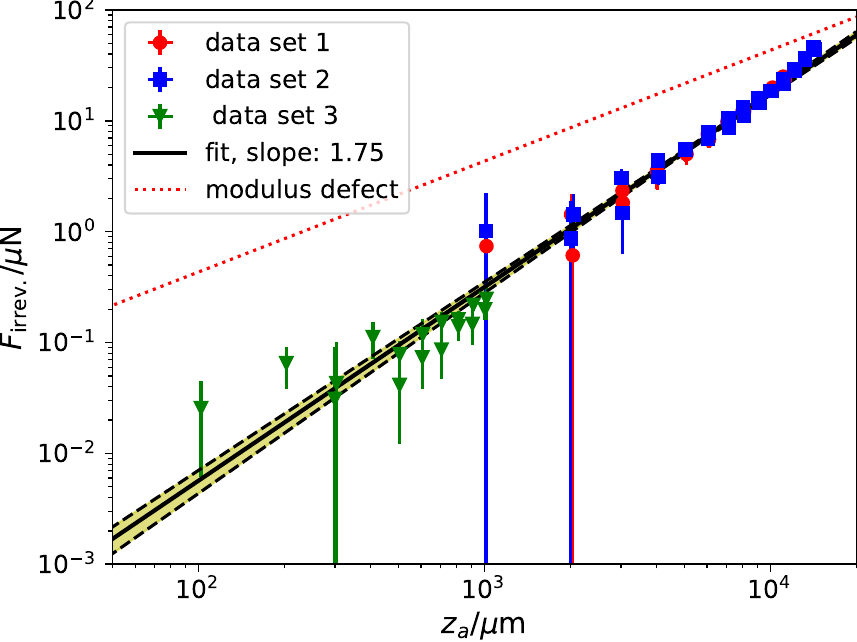}
\end{center}
\caption{Points with error bars denote the irreversible effect measured after an excursion of the balance given by the horizontal axis in \si{\micro\meter}. The different colors and symbols of the points denote three different data sets. Each excursion was held for \SI{30}{\second}. The solid black line is a power law fit to the data, and the shaded area gives the 1-$\sigma$ uncertainty band of the fit line. For comparison, the red dotted line gives the magnitude of the anelastic-after effect at $t^\ast=0$, which is for all excursions shown here much larger than the irreversible effect.
\label{fig:irreversible:effects}}
\end{figure}

At this point, we do not know the precise mechanism that causes the irreversible effects. Many causes are discussed in the literature, such as dislocation pinning~\cite{deSalvo2011}, work hardening, clamp effects, lattice-internal stick-slip effects, and more. 

Despite not knowing the physical mechanism,  we can measure the irreversible effect and discuss how much it affects the Kibble experiment. Fig.~\ref{fig:irreversible:effects} shows the results of three such measurements. For each measurement, the balance was held at a given excursion for \SI{30}{\second} and then controlled back to zero, where the force required to hold the balance at zero was recorded. The drift corrected force as a function of excursion is shown in Fig.~\ref{fig:irreversible:effects}. Three data sets were taken, two with larger excursions from \SI{1}{\milli\meter} to more than \SI{10}{\milli\meter} and one set with excursions from \SI{0.1}{\milli\meter} to \SI{1}{\milli\meter}. Besides a small discontinuity, between the set with smaller excursions and the two with larger excursions, caused probably by the drift subtractions, the data follow an exponential law, $F_\mathrm{irrev.}\propto z_a^\beta$ with an exponent of $\beta=1.73$. For a \SI{50}{\micro\meter} excursion, the irreversible effect contributes a force of $\SI{2}{\nano\newton}\pm\SI{1.5}{\nano\newton}$. The figure also shows, for comparison, the effect caused by the modulus defect, which is proportional to excursion; see the dotted red line. For small displacement, the anelastic effect immediately after bending, i.e., $t^\ast=0$ is much larger than the irreversible effect. A \SI{50}{\micro\meter} excursion causes an anelastic aftereffect of \SI{0.2}{\micro\newton} or about 100 times larger than the irreversible effect. Fortunately, as has been discussed in this article, the anelastic effect decays away and can be mitigated with an erasing procedure.

\section{Conclusion} 

The purpose of this article is to explore the usage and limitations of flexure-based Kibble balances with the hopes of breaking ground for the widespread usage of flexures in the Kibble balance, similar to what the seminal paper~\cite{Quinn1992} has done for conventional balances. To achieve this goal, we have reviewed the basic considerations necessary to understand the effect of anelastic relaxation on the Kibble balance measurement. With the Laplace transformation, it is possible to calculate the aftereffect of sinusoidal excitation on the stress produced by the flexure. To mitigate the anelastic aftereffect on the Kibble balance measurement, an erasing procedure can be employed after the velocity mode. The simplest erasing procedure is the boxcar eraser, holding the flexure at a specific deflection for a given time. A numerical calculation of the boxcar eraser was performed, and pairs of optimal deflection amplitude and hold time were found. 

The theoretical framework has been supported by measurements on a simple beam balance with a total of three flexures, one main and two end flexures. After straining the balance, the stress followed the relaxation predicted by theory. The application of a simple boxcar eraser significantly and consistent with the theory reduces the relaxation stress. After erasing, the largest excursion left an initial stress amplitude in the balance of approximately \SI{2.5}{\micro\newton}. This amplitude can be suppressed by a filter in the data analysis such as the one discussed in~\cite{swanson_2010}. It can reduce the effect of drift using realistic measurement times to \SI{2.4e-5}{} times the initial amplitude. Combining these two numbers leaves a systematic bias of \SI{0.36}{\nano\newton} for the Kibble balance experiment. This number is much smaller than the \SI{2}{\nano\newton} that were initially budgeted for the anelastic uncertainty for the Quantum Electro-Mechanical Metrology Suite, the new Kibble balance under construction at NIST.

From the measurements, a modulus defect of the hardened Copper Beryllium alloy C17200 with TF00 temper can be calculated. We find it to be $\eta=\SI{1.2e-4}{}\pm \SI{3e-5}{}$ slightly larger than the values reported in the literature previously.

While investigating the anelastic relaxation irreversible effects most prominent after large excursions were found. After a large excursion, the force required to servo the balance to the null position changed. Such effects are not unusual and were also seen previously in a Kibble balance with a knife edge, NIST-3~\cite{Schla2015} in agreement with a statement made in\cite{Quinn1992b}. At this point, we have not found a convincing explanation for the irreversible effect. A systematic investigation showed that the force change has a power law dependence on the excursion with an exponent of 1.73.

The irreversible effect after excursions in the velocity mode will be common to all weighings and should not cause a systematic bias. A possible systematic bias can occur by the irreversible effect caused by balance excursions that occur during mass exchanges. Since the excursions for the mass-on and mass-off measurements have different signs, a systematic bias will remain.

However, the excursion in mass exchange should be small for a well tuned Kibble balance, and the irreversible effect should be in the single digit nanonewton level,  as shown in Fig~\ref{fig:irreversible:effects}. Thus, is is not deemed to be a limit for accuracy for the present requirements to performance of the QEMMS.

Lastly, while thinking about these flexures and the physics that describes them, we had the idea of a magic flexure --- a flexure geometry that promises to have very little anelastic relaxation. Unfortunately, at this point, we do not have the resources to investigate the idea experimentally. A preliminary numerical investigation, however, implies that it is possible to build flexures with very low loss. The key idea is to shape the cross-section of the flexure such that the region of maximal bending moves to a thinner part as the elastic modulus increases and compensates for the increase in the elastic modulus.

Flexures are amazing elements for precision mechanics, and it is time for widespread adoption of them in the main and auxiliary pivots for Kibble balance. We firmly believe flexures are the best choice for pivots in high-precision mechanical experiments.

\section{Acknowledgement}
We would like to thank our colleagues Bill Luecke and Jon Pratt at the National Institute of Standards and Technology for their helpful comments on the manuscript.
\section*{References}


\providecommand{\newblock}{}

\end{document}